\documentclass{elsart}
\usepackage{graphicx}
\usepackage{amsmath,amssymb}
\usepackage{colordvi}

\newcommand{\be}{\begin{equation}}
\newcommand{\ee}{\end{equation}}
\newcommand{\ba}{\begin{eqnarray}}
\newcommand{\ea}{\end{eqnarray}}
\newcommand{\n}{\nonumber}

\begin{document}
\begin{frontmatter}
\title{Talbot effect in cylindrical waveguides}

\author[label1,label2]{L. Praxmeyer}
\author[label1,label3]{K. W\'odkiewicz}
\address[label1]{Institute of Theoretical Physics, Warsaw University,
 00--681 Warsaw, Poland}
\address[label2]{Theoretical Physics Division,
Sofia University, James Bourchier 5 blvd,
1164 Sofia, Bulgaria }
\address[label3]{Department of Physics and Astronomy, University of New Mexico,
Albuquerque NM 87131, USA}
\begin{abstract}
We extend the theory of  Talbot revivals for  planar or
rectangular geometry to the case of cylindrical waveguides. We
derive a list of conditions that are necessary to obtain revivals
in cylindrical waveguides. A phase space approach based on the
Wigner and  the Kirkwood-Rihaczek  functions provides a pictorial
representation of  TM modes interference associated with  the
 Talbot effect.
\end{abstract}
\end{frontmatter}

\section{Introduction}
Although the Talbot effect was discovered firstly in the beginning
of nineteen century (1836) and then rediscovered many times in
different systems, it has never been a widely known phenomenon. In
quantum systems it is often referred  as ``self-imaging" or
``quantum revivals". Even in the optical domain not always Talbot
name is mentioned when this phenomenon is described and discussed.
However, it was W. H. F. Talbot, who firstly observed that
monochromatic light passing through periodic grating at a certain
distance from the grating forms its ideal image and consecutively
at integer multiples of this distance similar images are
reproduced. He also demonstrated  that when  white light is used,
images of the grating of different colors are formed at different
distances and that is why we believe that all ``self-imaging"
effects should be referred as Talbot effects.

After the original paper by Talbot \cite{talbot0}, followed by the
work of Lord Rayleigh \cite{Ray}, and then by a series of papers
of Wolfke \cite{wolfke}, many  papers have been written about this
subject. A comprehensive description of the Talbot effect and its
rediscoveries  in classical optics can be found in Patorski
\cite{patorski}. Similar historical review and a detailed
description of ``quantum revivals" can be found in \cite{169}. In
the last years of the XX century fractional aspects of the Talbot
effect attracted considerable interest
\cite{berry1,frTal,berry2,qcar1}. These effects were studied both
in optical and quantum mechanical domains. As regards Talbot
effect in optical waveguides most significant are works of Ulrich
\cite{ulrich}, who investigated waveguides of planar and ribbon
geometries.

 The literature concerning the
Talbot effect in planar or rectangular waveguide geometries is
quite reach (even some US patents for applications of the Talbot
effect in those systems exist), but the much more practical
cylindrical geometry was not taken into account in optical
studies. We present in this paper  a  comprehensive theoretical
study of the Talbot effect in cylindrical waveguides. This should
fill the gap in the available descriptions of self-imagining
phenomena in various waveguides. There is a limited number of
papers mentioning cylindrical geometry in quantum revivals
\cite{Roh}, but the results presented in this paper go beyond the
one obtained so far.

The paper is organized as follows. In Section \ref{sec2} we
present general assumptions from which our study starts. We use  a
formal similarity between field propagation in waveguides and the
dynamics of wave packets in potential wells in a given geometry.
Using this analogy phase space  Wigner and   Kirkwood-Rihaczek
functions are used. The phase space functions provide  a pictorial
description of  interference effects, that are the basis of the
Talbot revivals.  In Section \ref{sec4}, the case of dielectric
fibers and the possibility of revivals in this most practical for
possible application system is studied. In  Section \ref{sec3},
solutions of the wave equation in cylindrical mirror waveguides
are analyzed  and requirements that have to be fulfilled to obtain
revivals are derived and imposed. A detailed study of
approximations used is included. Finally, a summary of the results
is given with a short paragraph devoted to the presentation of the
applied method.

\section{Talbot effect in phase space}\label{sec2}
\subsection{General assumptions}
The reason why the Talbot effect appears both in optical and
quantum mechanical systems, mathematically can be summarized  very
briefly: The Helmholtz equation is common for classical
electrodynamics and quantum mechanics. The dependence  of the
electromagnetic field in the direction of field propagation can be
regarded as an analogue of time dependence of a  wave packet in
quantum mechanics. In the regime where paraxial approximation is
justified this analogy is especially clear.

This paper is focused on Talbot revivals of an initial field in
cylindrical waveguides which  is purely an optical phenomenon but,
nevertheless,  as we shall see in the next paragraphs, its most
simple explanation can be given referring to quantum mechanical
concept of phase space distributions. The key question we shall
pose and answer is whether the Talbot effect in cylindrical
waveguides exists and can be used in practice.

We assume that harmonic, monochromatic plane waves propagate
through the waveguide which symmetry axis was chosen as the $z$
direction of the system. Inserting the fields
\begin{eqnarray}
{\mathbf{E}}(x,y,z,t)={\text{E}}(x,y)e^{\pm ikz-i\omega t},\qquad
{\mathbf{B}}(x,y,z,t)={\text{B}}(x,y)e^{\pm ikz-i\omega
   t} \,,\label{pola}
\end{eqnarray}
into the Maxwell equations,  we obtain the two-dimensional
Helmholtz equation:
\begin{displaymath}
\big[\nabla^2_{\perp}+\gamma^2
\big]\left(
\begin{array}{c}
{\text{E}}\\{\text{B}}
\end{array}
\right)=  0\,\n, \quad {\text{where}}\quad \gamma^2=
\mu\varepsilon\frac{\omega^2}{c^2}-k^2\,,\quad
\nabla^2_{\perp}=\nabla^2-\frac{\partial^2}{\partial z^2}\,.
\end{displaymath}

 The
propagating constants $k_i$ corresponding to specific modes are to
be derived from appropriate boundary conditions (see, e.g.
\cite{jackson}).

\subsection{Phase space distributions}
As we have already mentioned the basis of Talbot effect, i.e.
interference, is  especially  clearly seen in the phase space
description. The quasi--distribution functions used for the study
of electromagnetic fields, especially pulses, usually work in the
time--frequency domain. Here, a totally different approach was
chosen as all the fields taken into consideration are assumed to
be monochromatic and of a harmonic time--dependence. Thus, we
shall use a typical phase space known from mechanics constructed
from the position and momentum variables.

Some complications result from the fact that the electromagnetic
field requires a vector description. However, in many cases (e.g.
mirror waveguides) solutions of the Maxwell equations are
divided into $TM$ and $TE$ modes, i.e. are of type 
entirely determined by the $E_z$ or $B_z$ field component,
respectively. Thus, we can treat
 $E_z$  ($B_z$) component, that  fully describes the field, as a single
scalar function corresponding to the quantum mechanical wave
function of a potential well problem with analogous geometry.

\begin{figure}[hb]
\begin{center}
\includegraphics[scale=.59]{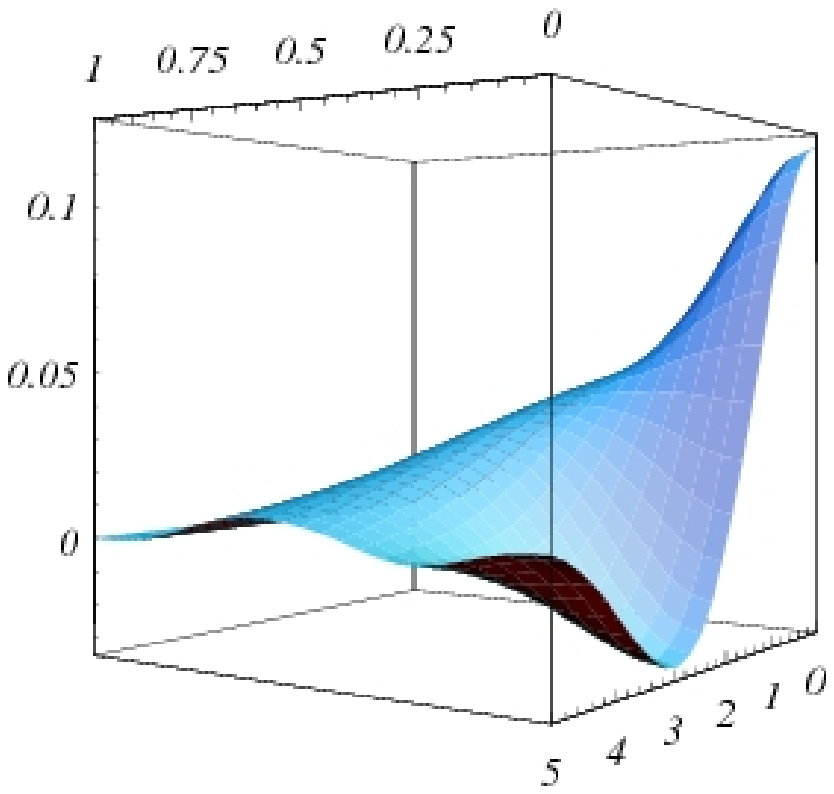}$\qquad$
\includegraphics[scale=.58]{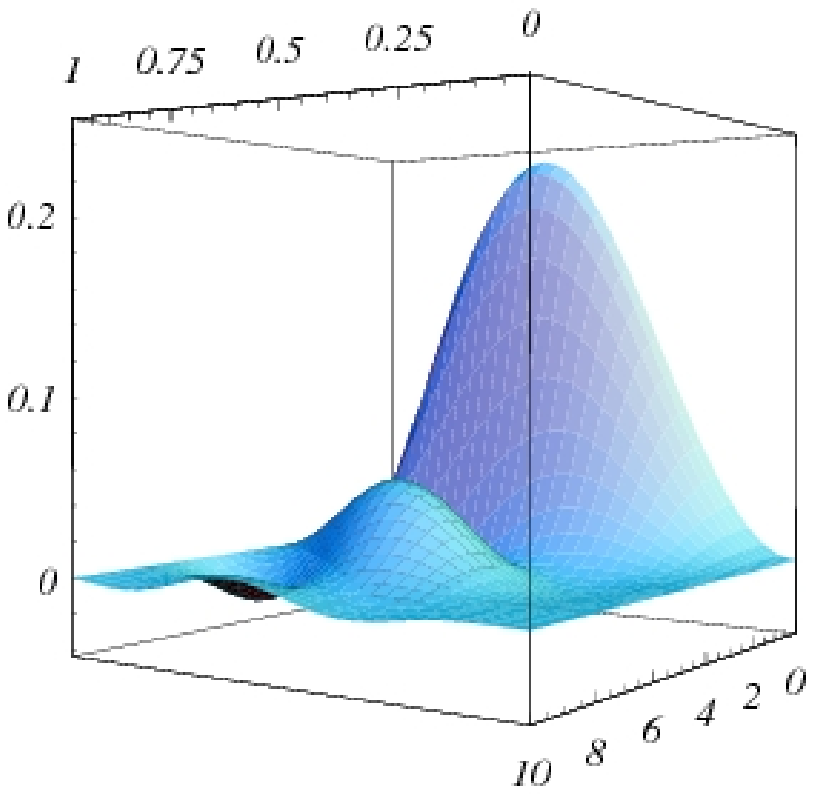}\\
\begin{picture}(0,0)(35,10)
\put(-140,30){\makebox(0,0){\footnotesize a)}}
\put(60,30){\makebox(0,0){\footnotesize b)}}
\put(6,28){\makebox(0,0){\footnotesize{$p_x$}}}
\put(-64,162){\makebox(0,0){\footnotesize{$x$}}}
\put(178,28){\makebox(0,0){\footnotesize{$p_x$}}}
\put(106,160){\makebox(0,0){\footnotesize{$x$}}}
\end{picture}
\end{center}
\caption{\footnotesize
The $y=0$ and $p_y=0$ cross-sections of ``transversal"
quasi--distributions corresponding to
$E_z=J_0(j_1 r)-J_0(j_2 r)$: a) cross-section of the Wigner function;
b)  cross-section of the real part of the K-R function.
 }\label{wig_bes2}
\end{figure}

To present problems of revivals is phase space we shall use the
following two quasi-distribution functions: the Wigner function
\cite{Wigner,Schleich},
\begin{equation}
W_{\Psi}(\vec{x},\vec{p})=
\frac{1}{(2\pi)^\mathrm{d}}\;\int
\Psi^{\star}(\vec{x}+\vec{\xi}/2)\; e^{{i\vec{p}\vec{\xi}}}\,
\Psi(\vec{x}-\vec{\xi}/2)\, d_\mathrm{d}\xi\;,
\label{zwig_powt}
\end{equation}
which is the most commonly known phase space distribution, and the
Kirkwood-Rihaczek (K-R) function \cite{Rih,K-R}:
\begin{equation}
K(\vec{x},\vec{p})=
\frac{1}{(2\pi)^\mathrm{d}} \int d_\mathrm{d}\xi\, \Psi^{\star}(\vec{\xi})\,
e^{{i(\vec{\xi}-\vec{x})\vec{p}}}\,\Psi(\vec{x}).
\label{zrih_powt}
\end{equation}
that is very convenient for calculations. Subscript
{\small$\mathrm{d}$} in the definitions above denotes number of
spatial dimensions of the system, whereas  $\Psi$ is a function
that characterizes a system (e.g. wave function in quantum
mechanical applications, or $E_z$ field component for TM modes in
waveguides, etc.) Because for 3 dimensional systems phase space
distributions are 6 dimensional, graphically only selected
cross-sections can be presented. This arbitrary choice is made
somehow easier  for solutions of the form of Eq. (\ref{pola}),
when the {\mbox{$z$-dependence}} of the field separates from the
transversal components.  Figure \ref{wig_bes2}  shows examples of
the ``transversal"  Wigner and K--R functions calculated for
$E_z=J_0(j_1 r)-J_0(j_2 r)$, which  corresponds to superposition
of TM$_{01}$ and TM$_{02}$ modes in cylindrical mirror waveguide:
$j_1$, $j_2$ denote the first and the second zero of Bessel
function $J_0(u)$ and $r=\sqrt{\frac{x^2+y^2}{a^2}}$, where $a$ is
a waveguide radius.

As we have already mentioned in the case of solutions of Maxwell
equations having the plane wave form we consider here, the
{\mbox{$z$-dependence}} of the field separates from the
transversal components. Thus, when describing Talbot effect, we
can simply ``forget" the transversal dependence of the field and
concentrate on the
$z$--$p_z$ cross-section of the phase space distribution which is
essential for the Talbot revivals. Integrals corresponding to the
{\small$\exp(-ik_i z)$} factors are quite elementary and the
cross-sections  of quasi-distribution functions for given $x$,
$y$, $p_x$, $p_y$, i.e. set the transversal position and momentum
components, are easy to obtain.

The Wigner function for a superposition of two plane waves
{\small$\exp(-ik_1 z)+\exp(-ik_2 z)$} is given by {\small
\ba
W_{\mathbf{2}}(z, p_z) \sim
\delta(p_z-k_1)+\delta(p_z-k_2)+
2\,\delta\left(p_z-\frac{k_1+k_2}{2}\right)\cos(z (k_2-k_1)).
\label{int_delta_2}
\ea
}The real part of the K--R distribution for such a superposition
takes the form {\small
\ba
\!\!\mathrm{Re}[K_{\mathbf{2}}(z,p_z)]\!\sim\delta(p_z\!-\!k_1)\!+
\!\delta(p_z\!-\!k_2)\!+\!
\left[\delta(p_z\!-\!k_1)\!+\!\delta(p_z\!-\!k_2)\right]\cos[z (k_2\!-\!k_1)].
\label{int_delta_2_rih}
\ea
}When there are $N$ superposed waves, the following sums are
obtained: {\small\ba W_{N}(z, p_z) \sim\sum_{i = 1}
\delta(p_z-k_i)+ 2\sum_{1 = i<j = N}
\delta
\bigg(p_z-\frac{k_i+k_j}{2}\bigg)
\cos[z (k_i-k_j)]\label{int_delta_n}
\ea
}for the Wigner function and, for the real part of the K--R
function, {\small
\ba
\mathrm{Re}[K_{N}(z, p_z)] \sim
\sum_{j = 1}^N \delta\left(p_z-k_j\right)
\bigg(1+\sum_{ i = 1,\, i\neq j}^N
\cos\left[ z (k_i-k_j)\right]\,\bigg),
\label{int_delta_n_rih}
\ea
}with appropriate coefficients. In both cases the whole $z$
dependence is inserted into interference {\small $\cos[z
(k_i-k_j)]$} terms. Initially, at {\small $z=0$}, all this cosines
are equal to 1. The further $z$ dependence is guided by {\small
$(k_i-k_j)$} factors. When all these {\small $(k_i-k_j)$} factors are
commensurable with each other, perfect regular revivals are
obtained at such
$z_{rev}$ for which all {\small $\cos[z_{rev} (k_i-k_j)]$} are
simultaneously equal to 1 again.
%

Obviously, all superpositions of just two different modes, like
Eqs. (\ref{int_delta_2}), (\ref{int_delta_2_rih}), will revive
perfectly at multiples of $\frac{2\pi}{k_2-k_1}$, no matter what
kind of waveguide we consider. But, when there is more superposed
modes, commensurability of all possible {\small $(k_i-k_j)$}
factors is needed to obtain perfect revivals. As the propagating
constants
$k_i$ vividly depend on the type of waveguide and its parameters,
such a commensurability is rather an exceptional then a typical
case. Even dealing with highly symmetric problems like  mirror
waveguides of  planar or square cross-sections we have to keep in
mind that  {\small $k_i=\sqrt{k_0^2 - \gamma_i^2}$} which means
that commensurability of $\gamma_i$'s (in these waveguides
$\gamma_i$'s are commensurable) is not automatically followed by
commensurability of propagating constants $k_i$. Only when linear
approximation of {\small $ \sqrt{k_0^2 - \gamma_i^2}$} holds,
commensurability of $\gamma_i$'s is sufficient and that is why we
shall often limit ourselves to the lowest from propagating modes.
A definite advantage of working within the range  where the linear
approximation of square root holds for systems having $\gamma_i$'s
proportional to each other is the fact that for a given wavelength
the Talbot distance is settled, it does not depend on superposed
modes. Otherwise, different modes superpositions shall revive at
different distances.

All the features characteristic for the Talbot effect can be
clearly explained by looking at interference
$\cos[z(k_i-k_j)]$ terms, their getting in and out of phase.
 The phase space
description brings us in a natural way to this simple idea and
indicates how fundamental this concept is.

\section{Talbot effect in dielectric waveguides}\label{sec4}
Cylindrical dielectric waveguides are called optical fibers. We
shall analyze only the
step--index fibers, i.e. the fibers with constant refractive
indexes in the core and the cladding, which are entirely
characterized by radii of core and cladding and their reflective
indexes
$n_1=\sqrt{\mu_1\varepsilon_1}$ and
$n_2=\sqrt{\mu_2\varepsilon_2}$.
Assuming that the cladding and the core differ only by dielectric
constants (magnetic permeabilities
$\mu_1=\mu_2$) the standard boundary conditions lead to the
following equation \cite{Cheo,skrypt} {\small
\ba
\left[\frac{1}{\Gamma}\frac{J'_\nu(\Gamma )}{J_\nu(\Gamma )}+
\frac{1}{\kappa}\frac{K'_\nu({\kappa})}{K_\nu({\kappa})}\right]
\left[\frac{n_1^2}{\Gamma}\frac{J'_\nu(\Gamma )}{J_\nu(\Gamma )}+
\frac{n_2^2}{\kappa}\frac{K'_\nu({\kappa})}{K_\nu({\kappa
})}\right]= \frac{\nu^2 \omega^2}{ k^2 c^2
}\left[\frac{n_1^2}{\Gamma^2}+
\frac{n_2^2}{{\kappa}^2}\right]^2 \,,
\label{fiber_general} \ea }where $a$ denotes the core radius,
$\,\,\Gamma=a\gamma\,$, $\,{\kappa}=a\beta\,$,
$\,\gamma^2=\mu_1\varepsilon_1\frac{\omega^2}{c^2}-k^2\,$ and
{\mbox{$\beta^2=k^2-\mu_2\varepsilon_2\frac{\omega^2}{c^2}$.}}
Although Eq. (\ref{fiber_general}) has a quite nice regular form
there is no way to solve it analytically. In the simplest case
when the field has no azimuthal dependence, i.e. $\nu=0$, its
solutions can be divided into  $TE$ and $TM$ type, as for the
mirror waveguides. Then a graphical picture gives a clear
representation of solutions similarly to  the case of finite
potential wells in quantum mechanics. However, we have to keep in
mind that the $\varphi-$independence, corresponding to $\nu=0$, is
not a typical case. General solutions of Eq. (\ref{fiber_general})
are $\varphi$ dependent
 and, actually, the lowest propagating mode in step-index dielectric
fiber is obtained for $\nu=1$. Thus, let us start our analysis
from general solutions.

\subsection{General solutions in step index fibers}
The lowest propagating mode in a cylindrical dielectric waveguide
is always an $HE_{11}$ mode (in this notation $HE$ means that
field $H_z$ dominates over $E_z$ field, while for $EH$ modes the
$E_z$ field dominates). Single mode fibers are of a great
practical importance, but in this study we are not interested in
them because a single mode propagates without a change in its
transverse distribution. Next modes are $TE_{01}$, $TM_{01}$, and
$HE_{21}$. They appear almost simultaneously as their propagating
constants are nearly the same.

Let us firstly consider a fiber in which there are only
these four propagating modes (e.g. $\lambda=1550 nm $, $n_1=1.46
$, $n_2=1.45  $, $a=4.5\mu m$). We shall refer to this situation as to the
``limit of small number of modes".   Numerical solutions of Eq.
(\ref{fiber_general}) obtained for these parameters are:
$a\gamma_{HE_{11}}=1.79268 $, $a\gamma_{TE_{01}}=2.75973 $,
$a\gamma_{TM_{01}}=2.76234 $, $a\gamma_{HE_{21}}=2.76342 $. While
analyzing problem of revivals of superpositions of  $HE_{11}$,
$TE_{01}$, $TM_{01}$, and $HE_{21}$ modes, one has to realize
that although  four modes are superposed, it is a special case
when three propagating constants (corresponding to
$TE_{01}$, $TM_{01}$, and $HE_{21}$ modes) are really close to
each other. As we have already mentioned,  superpositions of
arbitrary two modes with propagating constants $k_1$, $k_2$ shall
revive at all integer multiples of $z_T=\frac{2\pi}{k_1-k_2}$. It
is easy to calculate that initial images constructed from
superposition of pairs ($HE_{11}$, $TyE_{01}$), or ($HE_{11}$,
$TM_{01}$) or ($HE_{11}$, $HE_{21}$) shall revive at multiples of
$z^t_1=3,40787mm$, $z^t_2=3.3967mm  $, $z^t_3= 3.39214mm$,
respectively. Those values are so close to each other that images
constructed from all four modes shall revive at a mean value which
is
$\bar{z}_T=3.39891mm  $. In numerical simulations of field propagation
one can observe the first, second and even 20th or 50th Talbot
revival. Obviously, higher revivals are less accurate, and after
several tenths of faithful revivals they get out of phase -- but
then, after some propagating distance they get in phase again, and
again some faithful revivals are obtained. Then the situations of
getting in and out of phase appropriate cosines repeats almost
cyclicly.

\begin{figure}[h]
\begin{center}
$\!$\includegraphics[scale=.5]{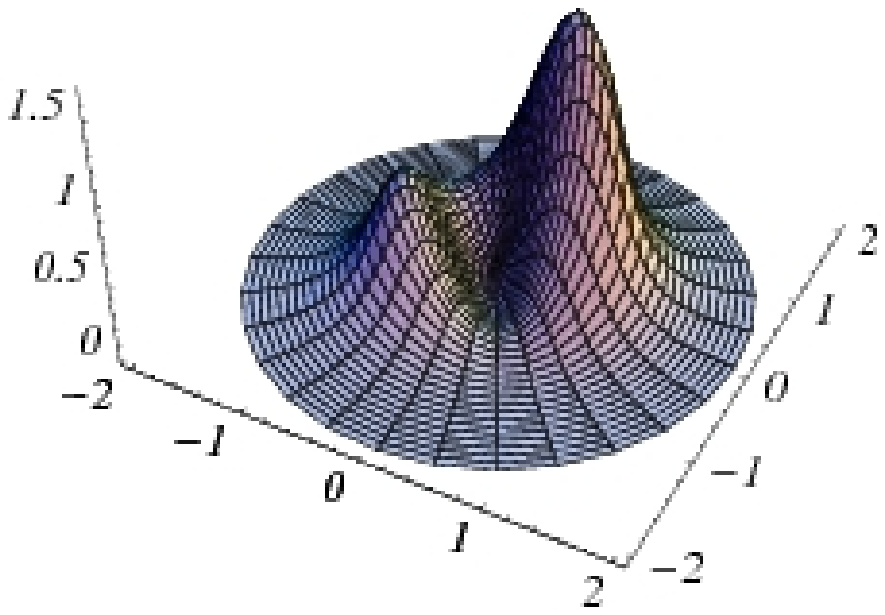}
\includegraphics[scale=.5]{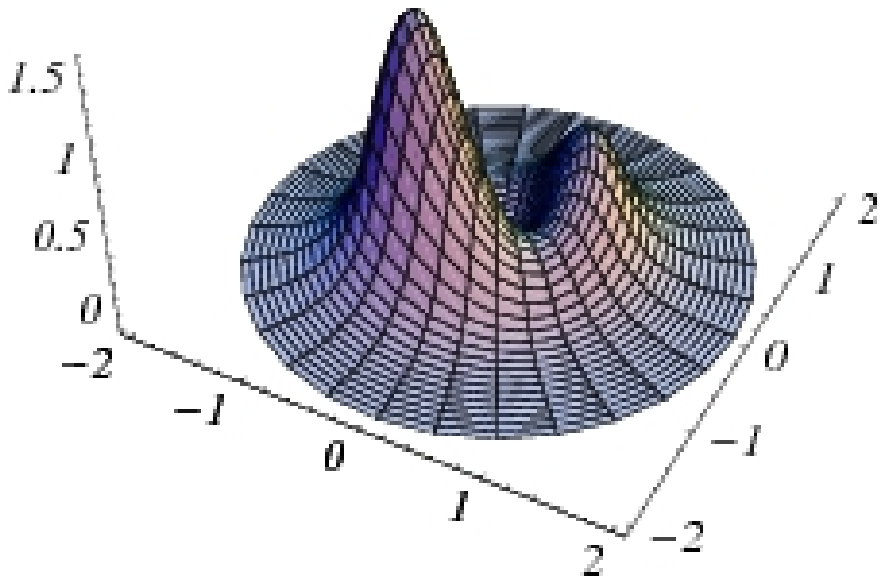}
\includegraphics[scale=.5]{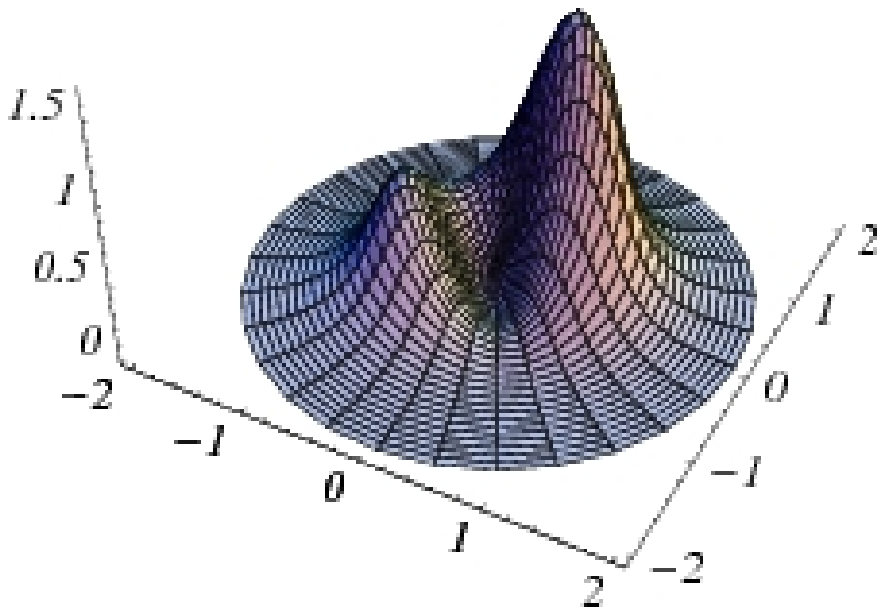}
\begin{picture}(0,0)(35,10)
\put(-350,24){\makebox(0,0){\footnotesize a)}}
\put(-220,24){\makebox(0,0){\footnotesize b)}}
\put(-90,24){\makebox(0,0){\footnotesize c)}}
\put(-60,104){\makebox(0,0){\scriptsize $f_0=$0.0000276}}
\end{picture}
\caption{\footnotesize Superposition of $TM_{01}-TE_{01}-HE_{11}-2HE_{21}$ modes:$\;$
a) the original intensity; b) intensity at the half of the Talbot
distance; c) 1st Talbot revival.
 }\label{bHE1_3D}\end{center}
\end{figure}

Plots of initial light intensity corresponding to superposition of
$TM_{01}-TE_{01}-HE_{11}-2HE_{21}$ modes, its first Talbot revival and a characteristic inverted
image at the half of the Talbot distance are presented in Figure
\ref{bHE1_3D}. Infidelities of subsequent revivals,
can be calculated using standard measure {\small
\ba
f_0(z)=\frac{\| I(0)- I(z)
\|}{\|I(0)\|}=\left(\frac{\int_0^{2\pi}d\varphi\int_0^1 d\rho\,
\rho\, |I(\rho,\varphi;0)-{I}(\rho,\varphi;z)|^2}{\int_0^{2\pi}d\varphi\int_0^1
d\rho\,
\rho\, |I(\rho,\varphi;0)|^2}\right)^{\frac{1}{2}},\label{measure}
\ea
}to which we refer as to the {\sl infidelity} as its value
increases with increasing deviation of
$I(z)$ from original
$I(0)$ and only when $f_0=0$ the copy is perfect.
For superposition of
$HE_{11}-TE_{01}-TM_{01}-2HE_{21}$ modes the infidelities of the
successive revivals are  quite low ($f_0=0.0000276$,
$f_0=0.0029825$, $f_0=0.0260536$, $f_0=0.0945819$ correspond to the 1st, 10th, 30th, and 60th
revival, respectively) and in such a four-mode fiber infidelities
of revivals for every initial field distribution would be of this
order.

It is seen that in the limit of small number of modes Talbot
revivals in dielectric fibers can be obtained. Change of fibers
parameters will result in the change of Talbot distance, but it is
quite obvious that for a small number of propagating modes Talbot
revivals can be obtained. Just as we have seen using phase space
representation -- appropriate number of cosines have to get in
phase  to obtain the revivals.

This simple and quite intuitive method of looking for the Talbot
distance starts to be more complicated when the number of
propagating modes increases. Then, different methods are have to
be applied to calculate revival distance as we shall see in the
next section on the example of $TE_{0i}$ modes.

\subsection{The $\varphi$--independent case ($\nu=0$)}
\noindent
The $\varphi$--independent solutions of Eq. (\ref{fiber_general})
can be divided into $TE$ and $TM$ modes.
 For $TE$ modes 
the following relation is obtained:
\be
\frac{1}{\Gamma}\frac{J_1(\Gamma )}{J_0(\Gamma )}+
\frac{1}{\kappa}\frac{K_1({\kappa})}{K_0({\kappa
})}=0\label{falK_eq}.
\ee
Because the parameters $\Gamma$ and $\kappa$ are (by definition)
correlated,
\be
\Gamma^2+\kappa^2=\frac{a^2 \omega^2}{c^2}
(n_1^2-n_2^2):=\mathrm{V}^2,
\label{falK_zamkn}
\ee
we obtain a set of equations that can be solved graphically or
numerically. It is convenient to introduce a normalized frequency
parameter V, Eq. (\ref{falK_zamkn}), and depict for example
$-\frac{1}{\Gamma}\frac{J_1(\Gamma )}{J_0(\Gamma )}$ and
$\frac{1}{\kappa}\frac{K_1({\kappa})}{K_0({\kappa })} $ as a
function of $ \Gamma$ for appropriate values  of V.

For $TM$ modes instead of Eq. (\ref{falK_eq}) we would have
\be
\frac{1}{\Gamma}\frac{J_1(\Gamma
)}{J_0(\Gamma )}+
\frac{\varepsilon_1}{\varepsilon_2}\frac{1}{\kappa}\frac{K_1({\kappa})}{K_0({\kappa })}=0.
\label{falK_eq2}
\ee
Graphical solutions of Eq. (\ref{falK_eq2}) are similar to those
of Eq. (\ref{falK_eq}), the only difference is that the Macdonald
part of the plot $\frac{K_1({\kappa})}{\kappa K_0({\kappa })}$ is
modified by a fixed factor
$\frac{\varepsilon_1}{\varepsilon_2}$, which is usually close to 1.
In practice this means that $TE_{0n}$ and $TM_{0n}$ modes have
nearly the same propagation constants and they will tend to appear
simultaneously.

In the previous subsection we were dealing with the small number
of modes, so now let us focus on a limit of large number of
propagating modes (large frequency parameter V). In the limit
$V\rightarrow\infty$, solutions $a\gamma_n$ would be given by
zeros of the Bessel $J_1(\rho)$ function, and propagating
constants would correspond exactly to those obtained for $TE_{0n}$
modes in the mirror waveguides. This case would be widely
discussed in the next Section, thus, now we shall take into
consideration only the finite values of V. Analyzing  the
graphical representation of Eq. (\ref{falK_eq}) for  different
values of frequency parameter V one finds that with the increase V
the Macdonald part of the plot, $\frac{K_1({\kappa})}{\kappa
K_0({\kappa })}$, starts to be parallel to $\Gamma$ axis for the
increasing range of
$\Gamma_n$'s. Moreover, it is also getting closer to this axis as
for $\Gamma\rightarrow 0$ value of $\frac{K_1({\kappa})}{\kappa
K_0({\kappa })}\simeq \frac{1}{V}$. For large V and low mode
numbers solutions would be of very regular form: the first one
corresponding to $TE_{01}$ mode will be given by, say
$\Gamma_0$, and the approximate formula for the next solutions
would be $\Gamma_n=\Gamma_0+ a n \pi$. Obviously the larger V, the
more accurate this formula is, and it  works well only for modes
having numbers low in comparison to the total number of modes.
Thus, in order to obtain revivals we shall use only a few percent
of the lowest from propagating modes  for constructing initial
images, namely those modes for which the linear approximation of
the square root is sufficient. Sometimes, however, it is easier to
omit such analytical approximations and to simply calculate
numerically infidelities of intensity distribution as a function
of $z$ and look for the minima of this function.

\subsubsection{Examples of revivals}
In this Section we consider only the $\varphi-$independent fields
that are fully characterized by their cross-section along the
radius. Thus, the figures presented shall show cross-sections of
the light intensity versus the normalized distance from a fiber
center
$\rho/a$. All the examples were calculated numerically for
a quite thick fiber ($a=1mm$) with refractive indexes of the core
and cladding equal to
$n_1=1.47$, $n_2=1.45$, respectively and a wavelength
of $\lambda=850 nm$.  For these values of  $\lambda$, $n_1$, and
$n_2$ the frequency parameter equals V$=1786.35$, which correspond
to more then 550 propagating $TE$ modes, and
$\gamma_0=3.82956/a$.

\begin{figure}[h]
\begin{center}
\begin{picture}(0,0)(35,10)
\put(40,14){\makebox(0,0){\footnotesize a)}}
\put(170,14){\makebox(0,0){\footnotesize b)}}
\put(40,-84){\makebox(0,0){\footnotesize d)}}
\put(300,14){\makebox(0,0){\footnotesize c)}}
\put(170,-84){\makebox(0,0){\footnotesize e)}}
\put(300,-84){\makebox(0,0){\footnotesize f)}}
\put(250,80){\makebox(0,0){\sl\scriptsize$f_0=8.52*10^{-6} $ }}
\put(380,80){\makebox(0,0){\sl\scriptsize $f_0=$0.00085 }}
\put(120,-16){\makebox(0,0){\sl\scriptsize$f_0=$0.00765 }}
\put(250,-16){\makebox(0,0){\sl\scriptsize$f_0=$0.03027 }}
\put(380,-16){\makebox(0,0){\sl\scriptsize$f_0=$0.08204 }}
\end{picture}
\includegraphics[scale=.53]{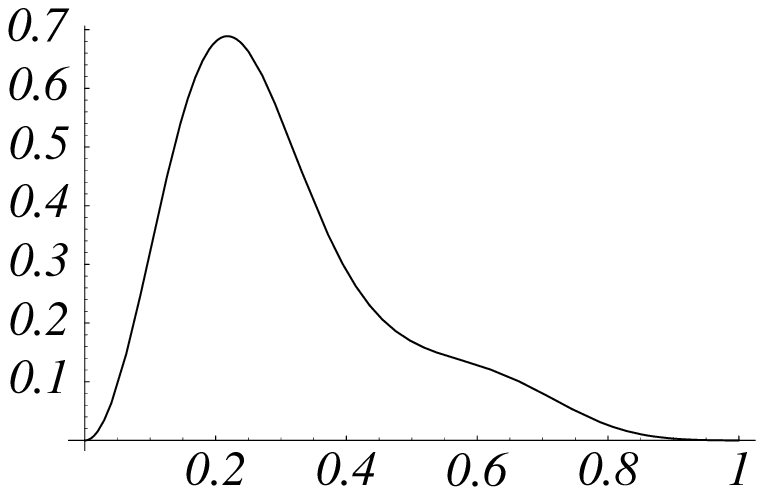}
\includegraphics[scale=.53]{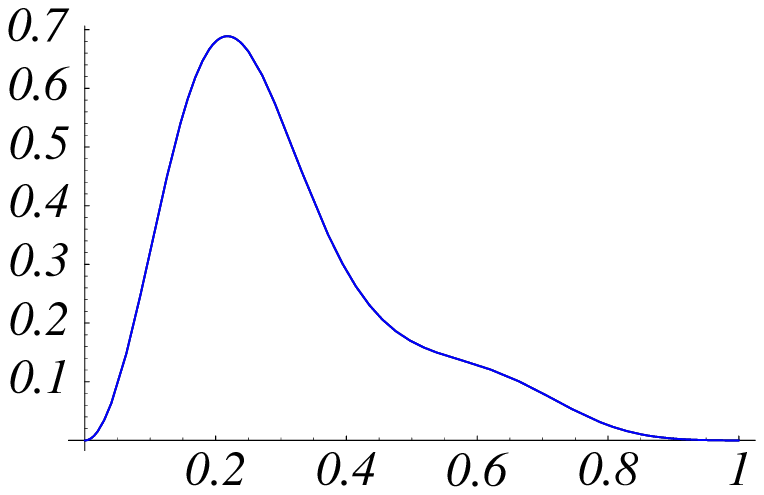}
\includegraphics[scale=.53]{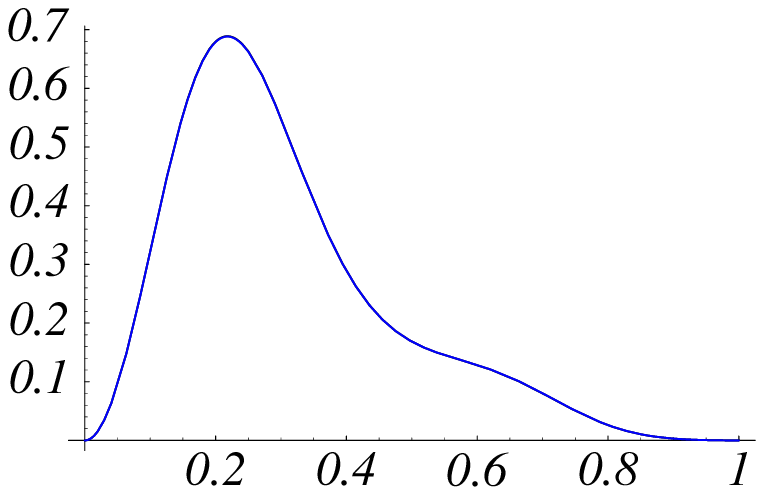}\\$\,$\\
\includegraphics[scale=.53]{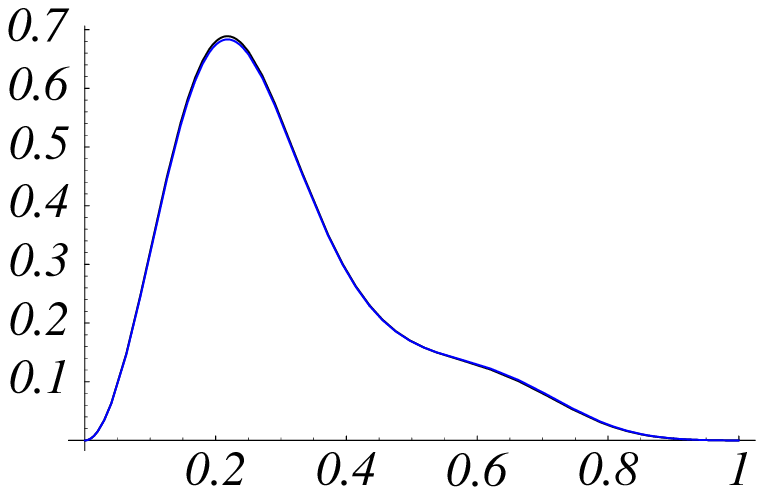}
\includegraphics[scale=.53]{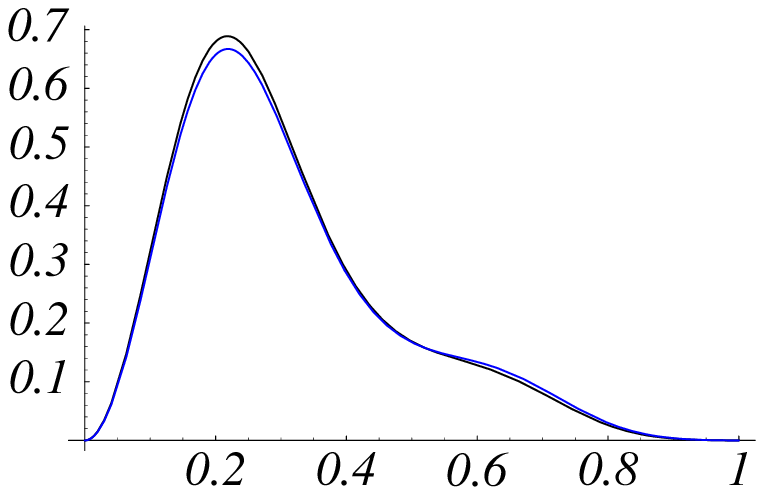}
\includegraphics[scale=.53]{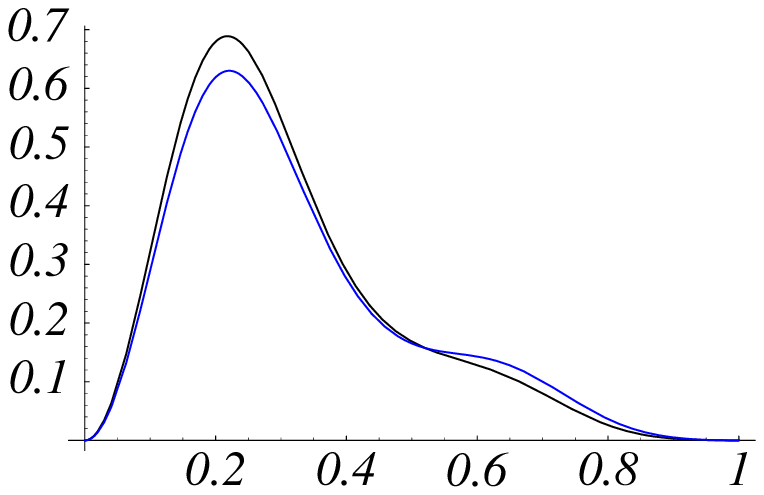}
\end{center}
\caption{\footnotesize Light intensity for the symmetric superposition
of $TE_{0n}$ for $n\in\{1,...,4\}$. The original intensity
distribution are plotted in black, the  intensity at the Talbot
distance $z_t=27.7038m$ and its multiples in blue. a) the original
intensity; b) 1st Talbot revival; {\mbox{c) 10th}} Talbot revival;
d) 30th Talbot revival; e) 60th Talbot revival; f) 100th Talbot
revival.
 }\label{a1mm4}
\end{figure}

\begin{figure}[h]
\begin{center}
\begin{picture}(0,0)(35,10)
\put(40,14){\makebox(0,0){\footnotesize a)}}
\put(170,14){\makebox(0,0){\footnotesize b)}}
\put(300,14){\makebox(0,0){\footnotesize c)}}
\put(120,80){\makebox(0,0){\sl\scriptsize$f_0=$0.7927 }}
\put(250,80){\makebox(0,0){\sl\scriptsize$f_0=$0.5907 }}
\put(380,80){\makebox(0,0){\sl\scriptsize $f_0=$1.0742 }}
\end{picture}
\includegraphics[scale=.53]{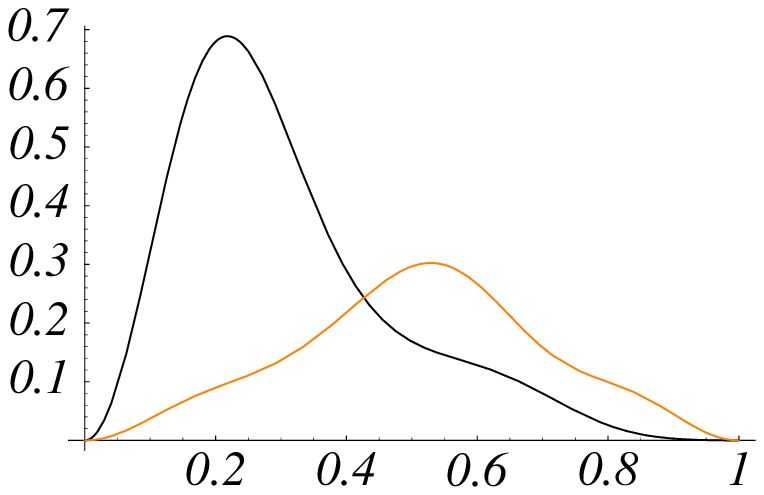}
\includegraphics[scale=.53]{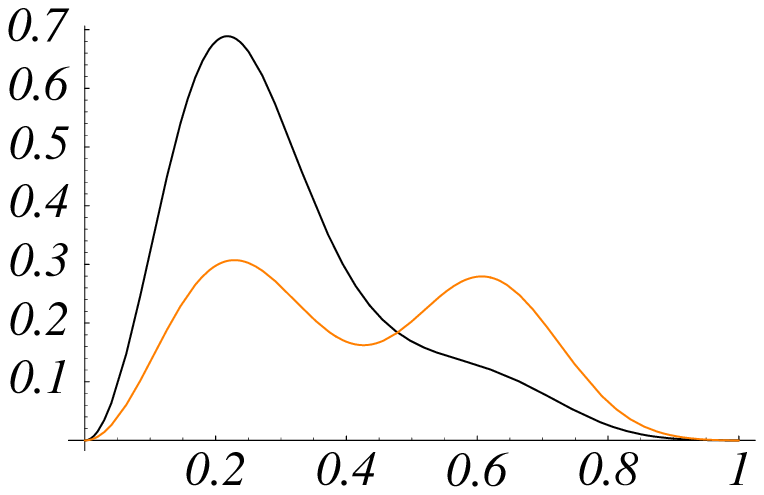}
\includegraphics[scale=.53]{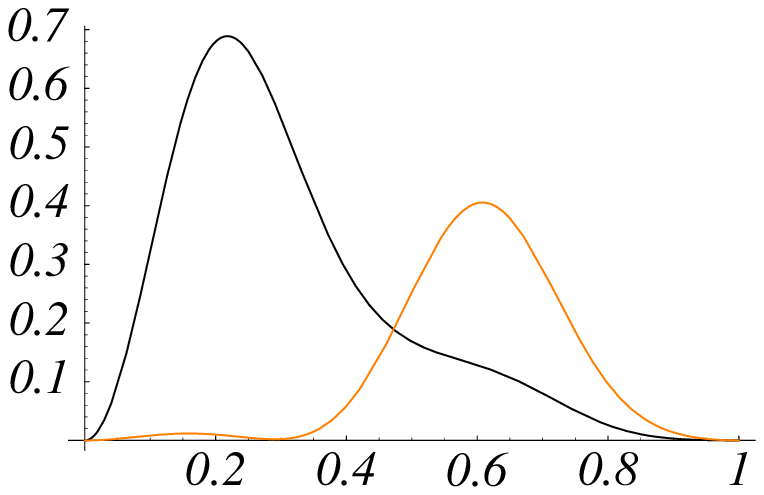}
\end{center}
\caption{\footnotesize The same superposition as in Fig. \ref{a1mm4}.
The original intensity distribution is plotted in black, in orange
are plotted intensities at a) $z=1m,\;$ b) $z=5m,\;$ c) $z=10m.\;$
 }\label{a1mm4pomiedzy}
\end{figure}

\paragraph*{{\footnotesize{a}})  symmetric superposition of
$TE_{01}+TE_{02}+TE_{03}+TE_{04}$ modes} $\,$\\
Figure \ref{a1mm4} presents the light intensity corresponding to
the superposition $TE_{01}+TE_{02}+TE_{03}+TE_{04}$ at $z=0$,
$z=z_t=27.7036m$,
 $z=10z_t$, $z=30z_t$, $z=60z_t$, and $z=100z_t$. The original intensity is
depicted in black, the revivals in blue. Above the plots of
revivals their infidelities are depicted. The value of the Talbot
distance $z_t$ was determined numerically by finding minimum of
the infidelity function, Eq. (\ref{measure}). It is seen that
although revivals are not perfect they are certainly faithful
enough even at distances of
$1km$. For comparison, Figure
\ref{a1mm4pomiedzy} presents examples of the light intensities at distances
between revivals: it is clear that the intensities at multiples of
the Talbot distance differ significantly from typical intensity
distribution during propagation.

\begin{figure}[h]
\begin{center}
\begin{picture}(0,0)(35,10)
\put(40,14){\makebox(0,0){\footnotesize a)}}
\put(170,14){\makebox(0,0){\footnotesize b)}}
\put(40,-84){\makebox(0,0){\footnotesize d)}}
\put(300,14){\makebox(0,0){\footnotesize c)}}
\put(170,-84){\makebox(0,0){\footnotesize e)}}
\put(300,-84){\makebox(0,0){\footnotesize f)}}
\put(255,90){\makebox(0,0){\scriptsize$f_0=8.4*10^{-6}$ }}
\put(385,90){\makebox(0,0){\sl\scriptsize $f_0=$0.00082}}
\put(125,-4){\makebox(0,0){\sl\scriptsize$f_0=$0.00730}}
\put(255,-4){\makebox(0,0){\sl\scriptsize$f_0=$0.02798}}
\put(385,-4){\makebox(0,0){\sl\scriptsize$f_0=$0.07252 }}
\end{picture}
\includegraphics[scale=.53]{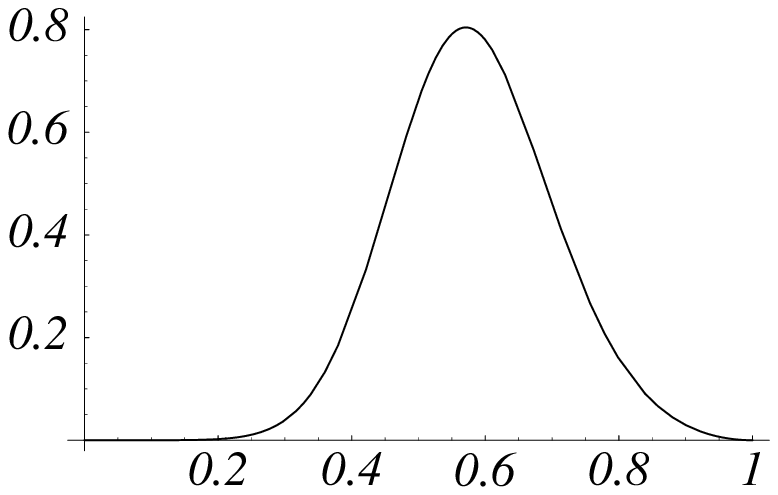}
\includegraphics[scale=.53]{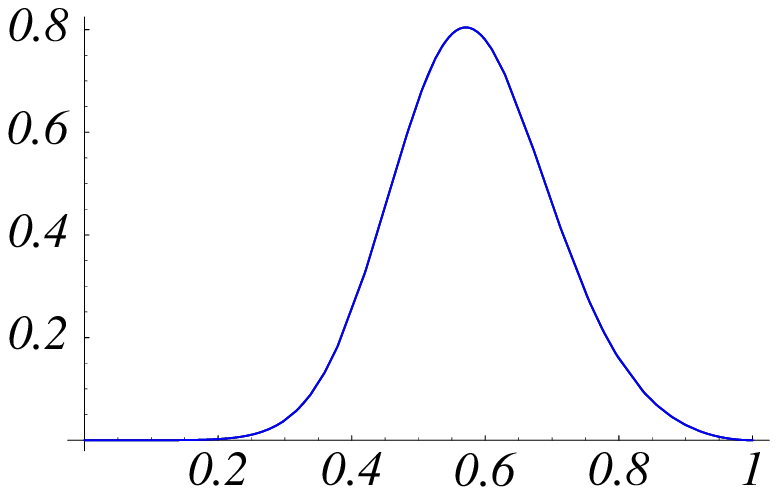}
\includegraphics[scale=.53]{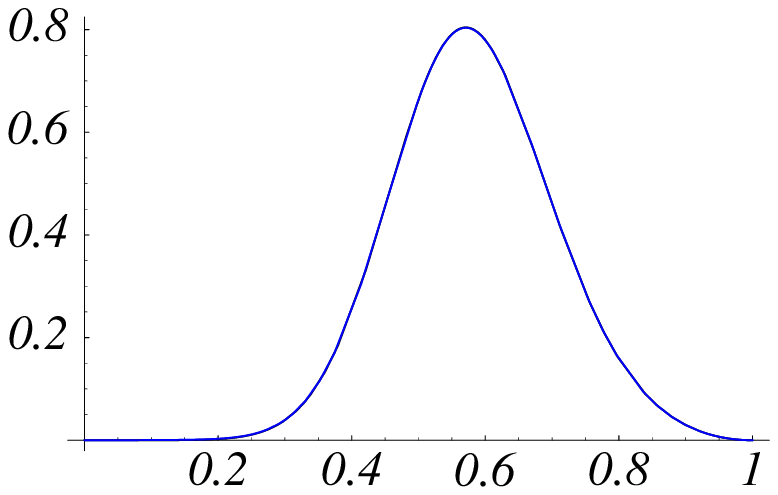}\\$\,$\\
\includegraphics[scale=.53]{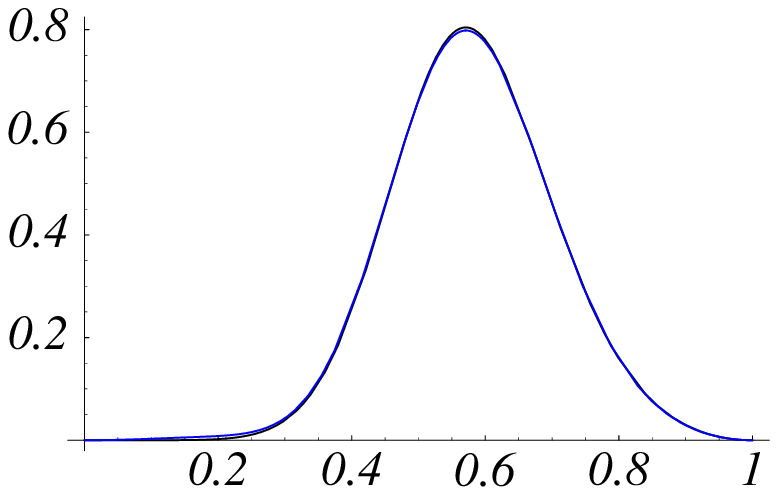}
\includegraphics[scale=.53]{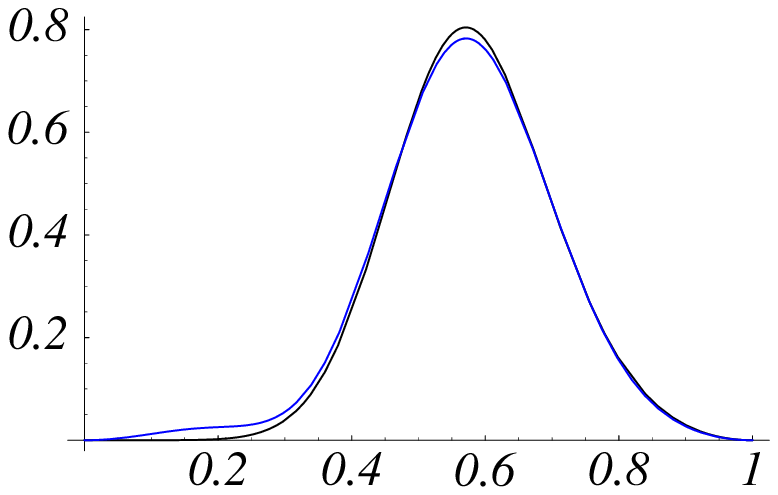}
\includegraphics[scale=.53]{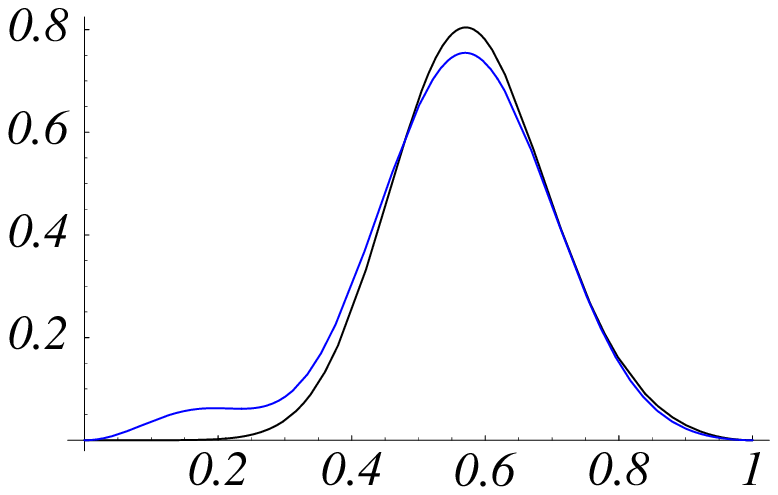}
\end{center}
\caption{\footnotesize Light intensity for the superposition
of $TE_{0n}$, $n\in\{1,...,25\}$, providing a Gaussian
distribution. The original intensity distribution are plotted in
black, the intensity at the Talbot distance $z_t=27.7035$ and its
multiples in blue. a) original intensity; b) 1st Talbot revival;
c) 10th Talbot revival; d) 30th Talbot revival; e) 60th Talbot
revival; f) 100th Talbot revival.
 }\label{a1mm25gauss}
\end{figure}

\paragraph*{{\footnotesize{b}}) Gaussian ``25"}$\,$\\
Figure \ref{a1mm25gauss} presents the revivals of the initial
intensity having a Gaussian radial distribution. The Gaussian
function was obtained from a superposition of the first 25
$TE_{0.}$ modes. The numerically calculated Talbot distance is
$z_t=27.7035m$ and the infidelities of revivals (depicted above
every plot) are quite low. It is clear that revivals of
intensities of a given shape can be observed. Obviously, this
result is true only on the assumption that other modes (e.g. modes
depending on $\varphi$) do not contribute to the initial image.
However,  the fidelities of revivals depend significantly on the
effective number of terms of the Bessel-Fourier (BF) series
contributing to the initial image. Although in the example
presented above we have taken first 25 terms of the BF series,
only first 9 coefficients were  larger then $1/1000$ and only
first 4 were larger then
$1/100$. The coefficients from 11th to 25 were of the order of
$5/10000$. If we would like to propagate, say, a more slim
Gaussian these proportions would be different. Is is worth noting
that in the case of dielectric waveguides this effective number of
contributing modes effect not only the fidelities of revivals, but
also the optimal Talbot distance, which we will clearly see in the
comparison with the next example.

\begin{figure}[ht]
\begin{center}
\begin{picture}(0,0)(35,10)
\put(40,14){\makebox(0,0){{\footnotesize a)}}}
\put(170,14){\makebox(0,0){{\footnotesize b)}}}
\put(40,-84){\makebox(0,0){{\footnotesize d)}}}
\put(300,14){\makebox(0,0){{\footnotesize c)}}}
\put(170,-84){\makebox(0,0){{\footnotesize e)}}}
\put(300,-84){\makebox(0,0){{\footnotesize f)}}}
\put(240,80){\makebox(0,0){{\sl\scriptsize$f_0=$0.00003 }}}
\put(370,80){\makebox(0,0){{\sl\scriptsize$f_0=$0.00274 }}}
\put(240,-14){\makebox(0,0){{\sl\scriptsize$f_0=$0.09218}}}
\put(110,-14){\makebox(0,0){{\sl\scriptsize$f_0=$0.02424 }}}
\put(370,-14){\makebox(0,0){{\sl\scriptsize$f_0=$0.23059}}}
\end{picture}
\includegraphics[scale=.53]{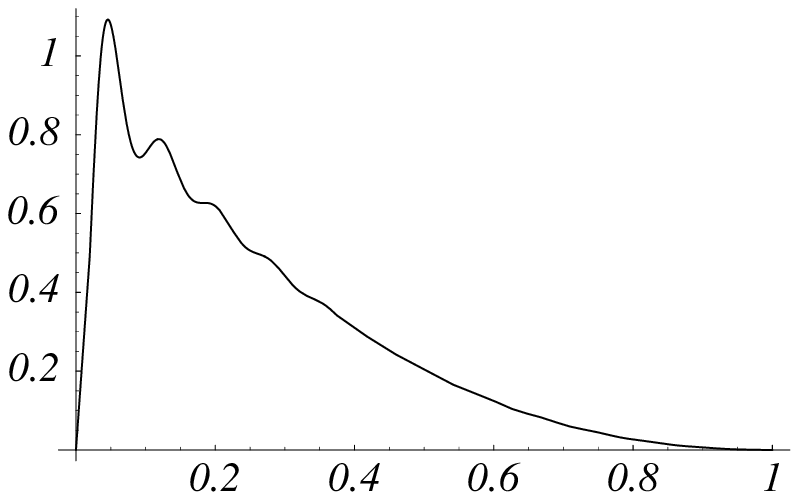}
\includegraphics[scale=.53]{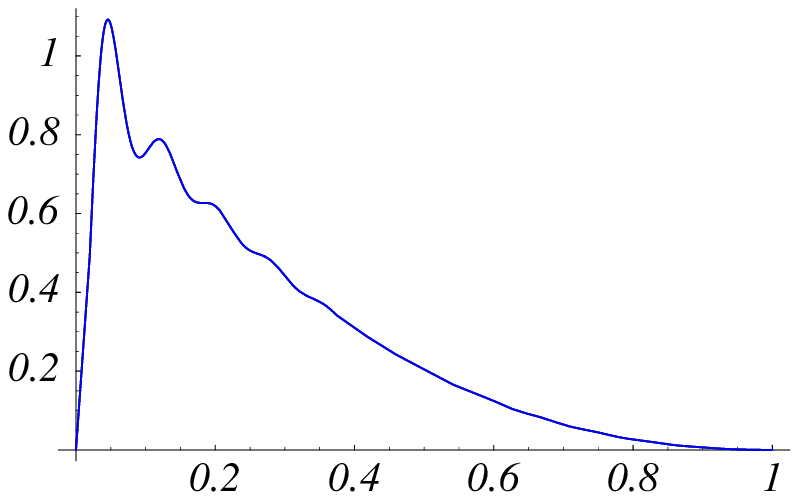}
\includegraphics[scale=.53]{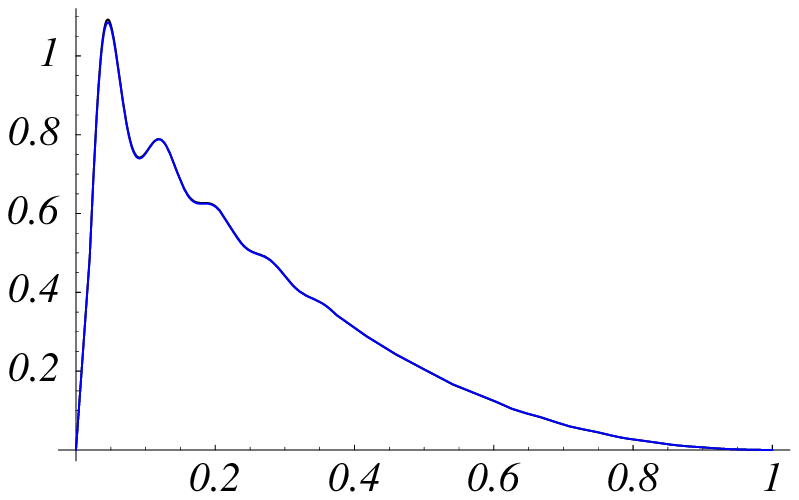}\\$\,$\\
\includegraphics[scale=.53]{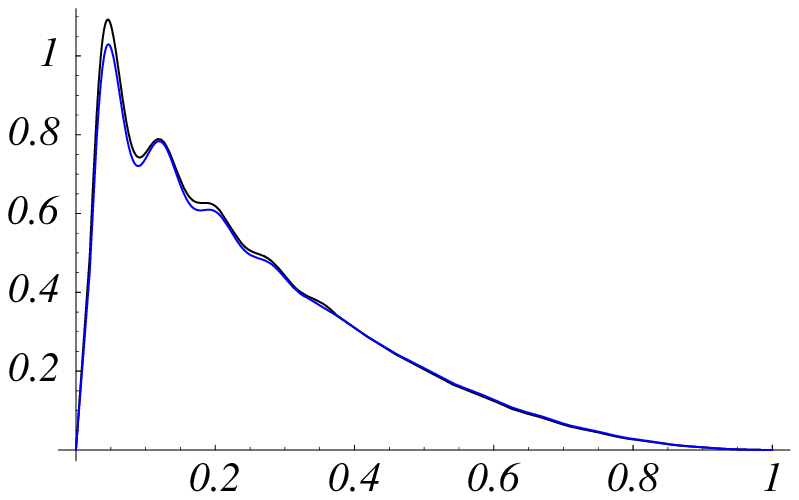}
\includegraphics[scale=.53]{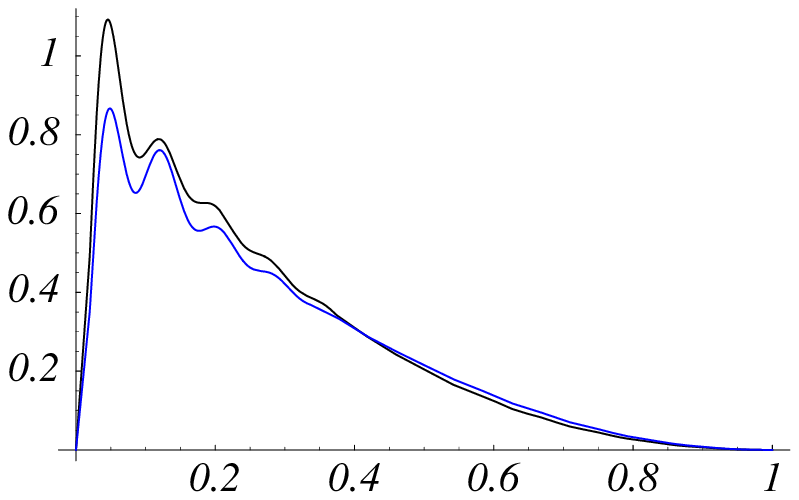}
\includegraphics[scale=.53]{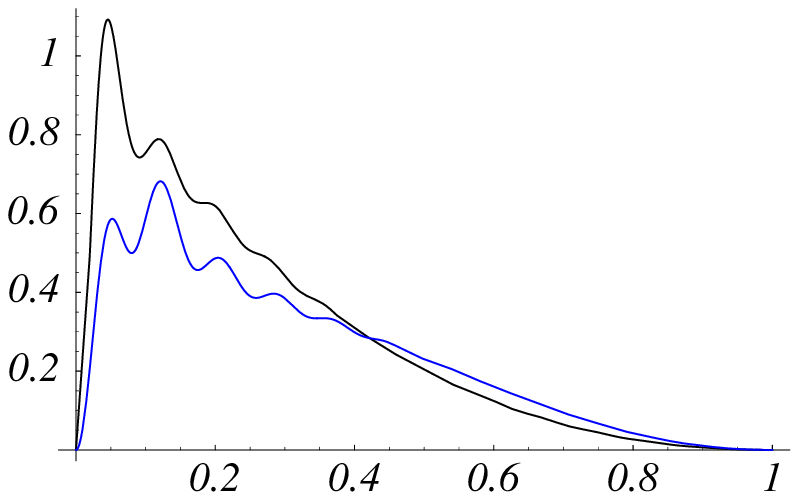}
\end{center}
\caption{\footnotesize
Light intensity for the symmetric superposition of  $TE_{0n}$
modes for
$n\in\{1,...,25\}$.
The original intensity distribution are plotted in black, the
intensity at the Talbot distance $z_t=27.7013m$ and its multiples
in blue. a) the original intensity; b) 1st Talbot revival; c) 10th
Talbot revival; d) 30th Talbot revival; e) 60th Talbot revival; f)
100th Talbot revival.
 }\label{a1mm25}
\end{figure}

\paragraph*{{\footnotesize c}) symmetric superposition of
$TE_{01}+TE_{02}+\dots+TE_{025}$ modes}$\,$\\
This example of revivals in a thick fiber illustrates how the
effective
 number of contributing modes might modify a Talbot distance. As the initial
intensity we take the one corresponding to the symmetric
superposition of $TE_{01}+TE_{02}+
\dots+TE_{025}$ modes.
Figure \ref{a1mm25} presents the  initial intensity, its 1st
Talbot revival and then 10th, 30th, 60th, and 100th Talbot
revival. The numerically calculated ``optimal" Talbot distance is
equal in this case  $z_t=27.7013m$, which is slightly smaller then
in the case presented in examples a), b).

We have already stressed the fact that different initial images
have different Talbot distances  can be explained by the effective
number of modes that are superposed. It is, again, a consequence
of importance of the quality of approximations used. In our
simplified analysis we have assumed that
$\gamma_n\simeq \gamma_0+n\pi$. On the one hand this is
true only for $n$ small enough for  $\frac{K_1({\kappa})}{\kappa
K_0({\kappa })}$ plot to be parallel to the  $a\gamma$ axis, on
the other  distance between neighboring zeros of Bessel functions
is $\sim \pi$ only for higher modes. Thus, it is obvious that if
only first few modes contribute to the initial image (or their
contribution dominates) the real Talbot distance would be slightly
different then if there were more contributing modes. Differences
in optimal Talbot distance $z_t$ for superpositions presented in
examples {\footnotesize a}), {\footnotesize b}), {\footnotesize c}) can be 
evaluated explicitly by calculating the infidelities of revivals.
They start to be important at 20th, or 50th revival when the
initial $2mm$ difference in optimal Talbot distances results in
$5$ or $10cm$ divergence from the distance of optimal revival.
As long as  we are interested only at the first Talbot revival we
can take an average $\bar{z}_t$ and the infidelities of revivals
of different initial images at $\bar{z}_t$ should not be larger
then 1/100 which is sufficient for most applications. 
Obviously, when we are interested in revivals at
larger distances all the initial images can be divided in classes
having the same  ``effective number of contributing modes" and the
revivals at multiples of the corresponding ``optimal Talbot
distances" would be obtained for all the images within the class.



We have shown numerical simulations indicating that the Talbot
revivals of initial images constructed from
$TE_{0n}$ modes can be obtained in optical fibers.
Although examples {\footnotesize a}) - {\footnotesize c}) present
superposition of $TE$ modes, one can note that for
$TM_{0n}$ modes revivals should be even more faithful, as
$\frac{\varepsilon_1}{\varepsilon_2}<1$ and thus $\gamma_n$ are
even closer to zeros of $J_1(\rho)$ function then it is for TE
modes, 
which brings us to the case of mirror waveguides.

\section{Talbot effect in mirror waveguides}\label{sec3}
Although we present it in the last Section, a model of ideal
mirror waveguides is very useful for preliminary calculations. It
is analytically soluble for systems of standard geometries because
boundary conditions are quite simple: normal component of
$\mathbf{B}$ and tangential component of $\mathbf{E}$ have to
vanish at the boundary mirror surface.

\subsection{Planar mirror waveguides}$\,$\\
In the elementary case of planar mirror waveguide 
 the propagation constant
for $n$th mode is given by $k_n^2=k_0^2-\frac{n^2 \pi^2}{d^2}$,
where $d$ denotes a separation distance between mirrors plates
\cite{fiber,Cheo}. It is clear that, in general, the field
changes its transverse
distribution as it travels through the waveguide because different
modes travel with different propagation constants and different
group velocities. The following expansion of the propagation
constant $k_n$,
\be
k_n=\sqrt{k_0^2-\frac{n^2\pi^2}{d^2}}=
k_0\sqrt{1-\frac{n^2\pi^2}{d^2 k_0^2}}
\simeq k_0\left(1-\frac{1}{2}\frac{n^2 \pi^2}{d^2 k_0^2}\right),\label{aprox}
\ee
shows, however, that within this approximation for $z=\frac{4 k_0
d^2}{\pi}$ the initial field is obtained. Obviously, requirements
for above linear approximation are not met for an arbitrary $k_n$.
Higher modes have to be prevented from contributing to the image,
because only then approximation of the square with accuracy to the
linear term is sufficient. The reason why revivals appear is that
$\gamma_i$ factors are all of the form {\sl constant
(characterizing the system)} times {\sl integer}. The question
arises whether in cylindrical mirror waveguides similar analytical
formula for the Talbot distance can be obtained.

Let us note here that necessity of taking care of paraxial
approximation is the main difference  between optical and quantum
mechanical Talbot revivals. In the case of infinite potential well
in quantum mechanics, eigenvalues of the system are given by
{\small $\frac{n^2\pi^2}{d^2}$} and no excluding of modes (wave
functions) that follows from linearization of square root is
needed.

\subsection{Cylindrical mirror waveguides}$\,$\\
Solutions of the wave equation in cylindrical mirror waveguides
corresponding to angular dependance
$e^{\pm i\nu\varphi}$, where $\nu$ denotes a number of  Bessel
function of a radial solution. Propagating constants are of the
form of {\small $\sqrt{{k_0^2}-\frac{j_{\nu n}^2}{a^2}}$} for TM
modes and {\small
$\sqrt{{k_0^2}-\frac{{j'_{\nu n}}^2}{a^2}}$} for TE modes, where
$j_{\nu n}$ denotes the $n$th root of  Bessel function
$J_{\nu}(x)$ and $j'_{\nu n}$ the $n$th root of its
derivative and $a$ is a waveguide radius.

In this case not only a linear approximation of square root is
needed but also approximation for zeros of Bessel function $j_{\nu
n}$ or $j'_{\nu n}$. Standard asymptotical expansions for $j_{\nu
n}$, $j'_{\nu n}$ are given by $j_{\nu n}\simeq
n\pi+\left(\nu-\frac{1}{2}\right)\frac{\pi}{2}$ and $ j_{\nu
n}'\simeq n\pi+\left(\nu-\frac{3}{2}\right)\frac{\pi}{2} $. They
are believed to be good enough for $n>\nu$ (or in more rigorous
manner for $n>2\nu$). Using these formulas we can repeat procedure
from Eq. (\ref{aprox}) and obtain: {\small{
\be
\sqrt{k_0^2-\frac{j_{\nu n}^2}{a^2}}=k_0\sqrt{1-\frac{j_{\nu
n}^2}{k_0^2 a^2}}\approx k_0
\left(1-\frac{1}{2}\frac{j_{\nu n}^2}{k_0^2
a^2}\right)\approx k_0-\frac{\pi^2(4n+2\nu-1)^2}{32 a^2 k_0}.
\label{przyblizenie}
\ee
}}Thus,
\begin{align*}
e^{i z\sqrt{{k_0^2}-\frac{j_{\nu n}^2}{a^2}}} \approx
    e^{ik_0z}e^{-i 2 \pi \left(\frac{\lambda}{128
    a^2}\right)(4n+2\nu-1)^2 z}
\end{align*}
and {\footnotesize
\be
(4n+2\nu-1)^2=
8\biggl(2n^2+2n\nu-n+\frac{\nu(\nu-1)}{2}\biggr)+1\n.
\ee
}Omitting common phase factor {\small $\exp\big({ik_0z+i
\frac{2 \pi\lambda}{128
    a^2} z}\big) $} we find that for a given
wavelength $\lambda$ and  waveguide radius $a$ at distance
 $z_T={16  a^2}/{\lambda} $ and its integer multiples
Talbot revivals are to be obtained. Similarly, for $TM$ modes
approximation for $j'_{\nu n}$ leads to the same Talbot distance
$z_T={16  a^2}/{\lambda}$.
\subsubsection{Examples of revivals in cylindrical mirror waveguides}
\begin{figure}[h]
\begin{center}
\begin{picture}(0,0)(35,10)
\put(135,80){\makebox(0,0){ {\sl\scriptsize $f_0^a=$ 0.02059}  }}
\put(300,80){\makebox(0,0){ {\sl\scriptsize $f_0=$ 0.07133}  }}
\put(75,-17){\makebox(0,0){ {\sl\scriptsize $f_0=$ 0.08621}  }}
\put(200,-17){\makebox(0,0){ {\sl\scriptsize $f_0=$ 0.14779}  }}
\put(330,-17){\makebox(0,0){ {\sl\scriptsize $f_0=$ 0.1429}  }}
\put(50,4){\makebox(0,0){\small $a)$}}
\put(184,4){\makebox(0,0){\small $b)$}}
\put(-15,-82){\makebox(0,0){\small$c)$}}
\put(120,-82){\makebox(0,0){\small$d)$}}
\put(248,-82){\makebox(0,0){\small$e)$}}
\put(15,80){\makebox(0,0){\small $ak=10^3$}}
\end{picture}
\includegraphics[scale=.55]{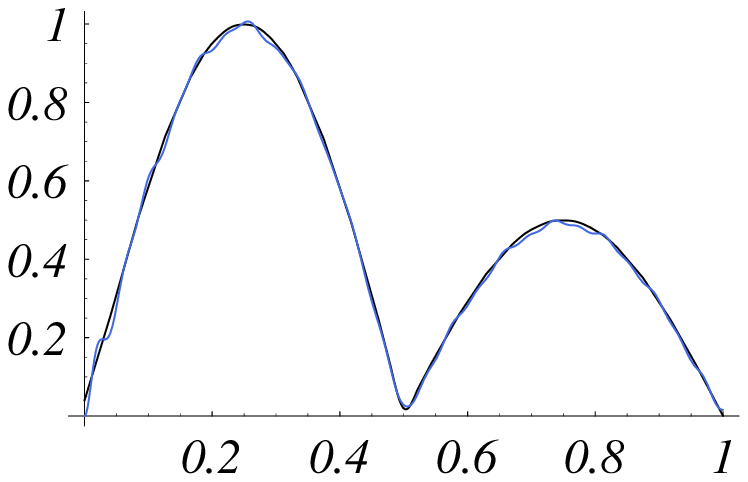}
\includegraphics[scale=.55]{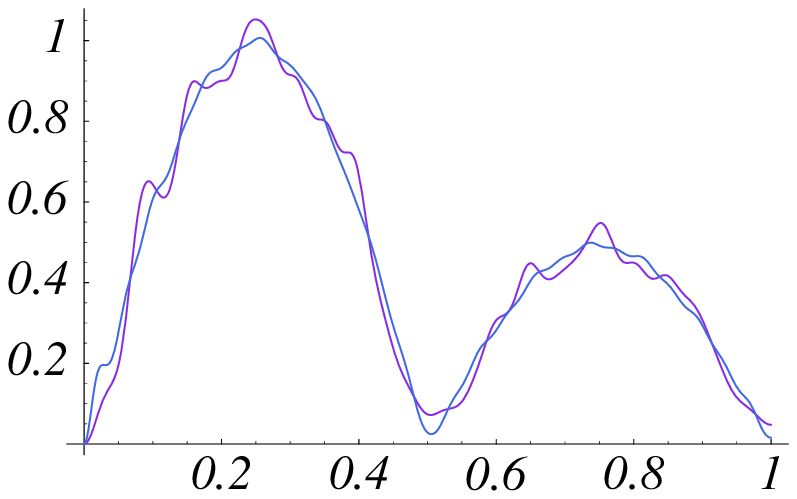}\\$\,$\\$\,$
\includegraphics[scale=.55]{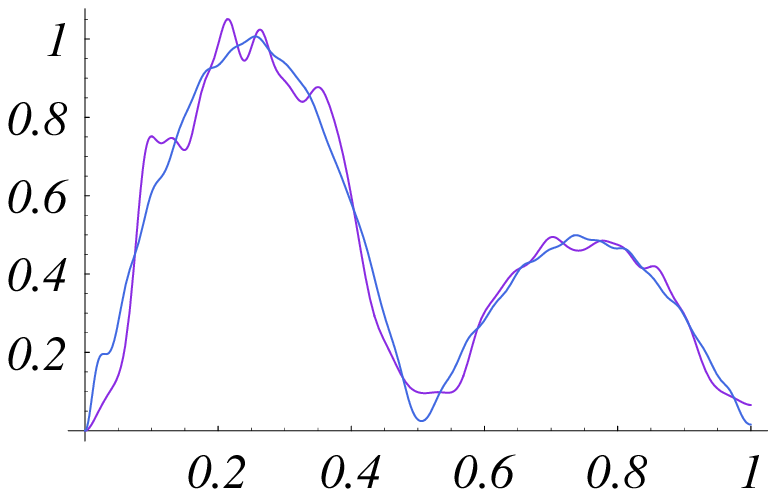}
\includegraphics[scale=.55]{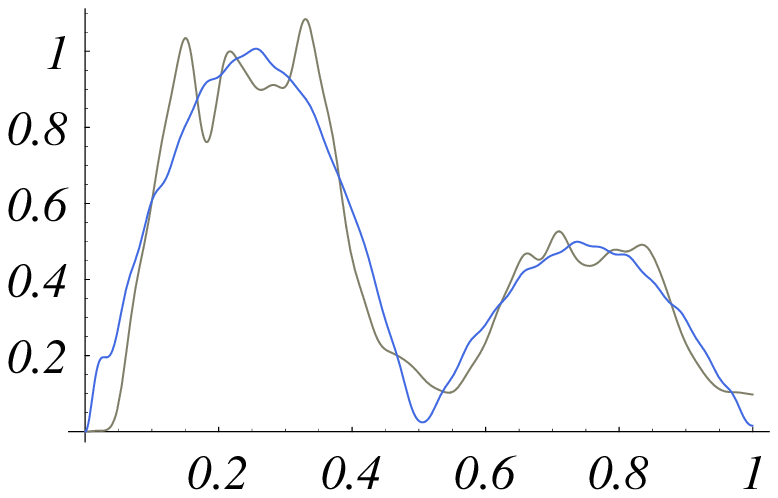}
\includegraphics[scale=.55]{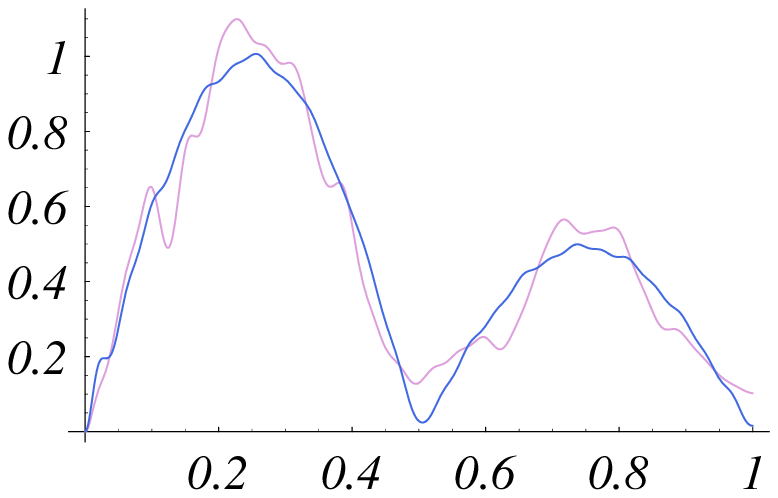}
\end{center}
\caption{\footnotesize {\underline{ $ I(\rho)=
\sin(2\pi \rho)$ for $\rho \in [0,\,0.5)$ and $I(\rho)=\frac{1}{2}
\sin(2\pi \rho)$ for $\rho \in [0.5,\,1]$}}.
The intensity of light is plotted: $\,a)$  shows initial
$I(\rho)$ (an origin and an approximation  given by the
first 50 terms of BF series, plotted together);
$\,b)$ shows the first Talbot revival at $z=z_T$ and the
initial intensity  shown together;
$\,c)$, $\,d)$, $\,e)$ show  2nd, 5th, and 10th Talbot revival,
respectively. }\label{sinn}
\end{figure}

\begin{figure}[h]
\begin{center}
\begin{picture}(0,0)(35,10)
\put(135,80){\makebox(0,0){ {\sl\scriptsize $f_0^a=$ 0.01657}  }}
\put(260,80){\makebox(0,0){ {\sl\scriptsize $f_0=$ 0.00511 }  }}
\put(75,-17){\makebox(0,0){ {\sl\scriptsize $f_0=$ 0.01252}  }}
\put(200,-17){\makebox(0,0){ {\sl\scriptsize $f_0=$ 0.01887}  }}
\put(330,-17){\makebox(0,0){ {\sl\scriptsize $f_0=$ 0.03586}  }}
\put(50,4){\makebox(0,0){\small$a)$}}
\put(184,4){\makebox(0,0){\small$b)$}}
\put(-15,-82){\makebox(0,0){\small$c)$}}
\put(120,-82){\makebox(0,0){\small$d)$}}
\put(248,-82){\makebox(0,0){\small$e)$}}
\put(15,90){\makebox(0,0){\small $ak=10^4$}}
\end{picture}
\includegraphics[scale=.55]{sinn50_org.eps}
\includegraphics[scale=.55]{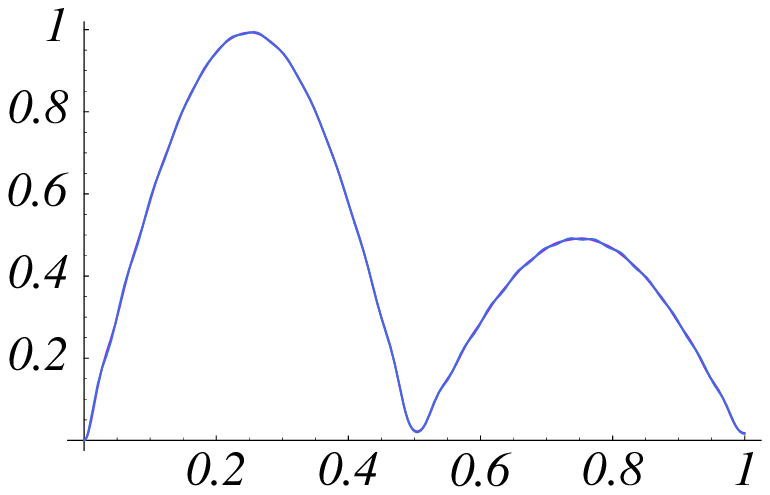}\\$\,$\\$\,$
\includegraphics[scale=.55]{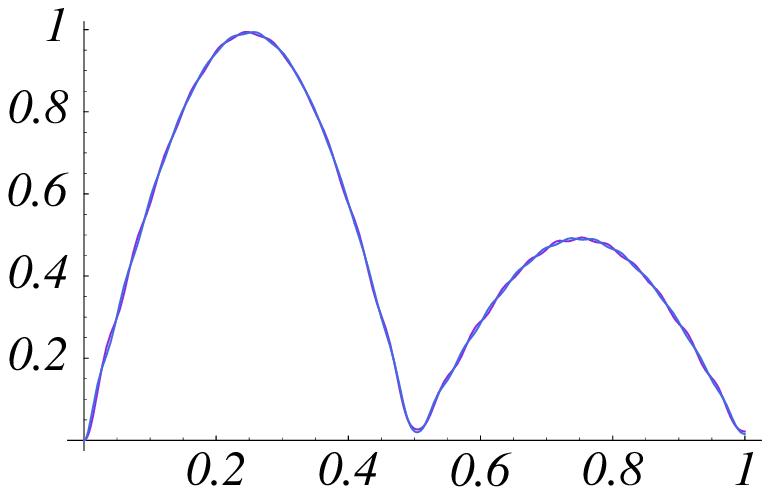}
\includegraphics[scale=.55]{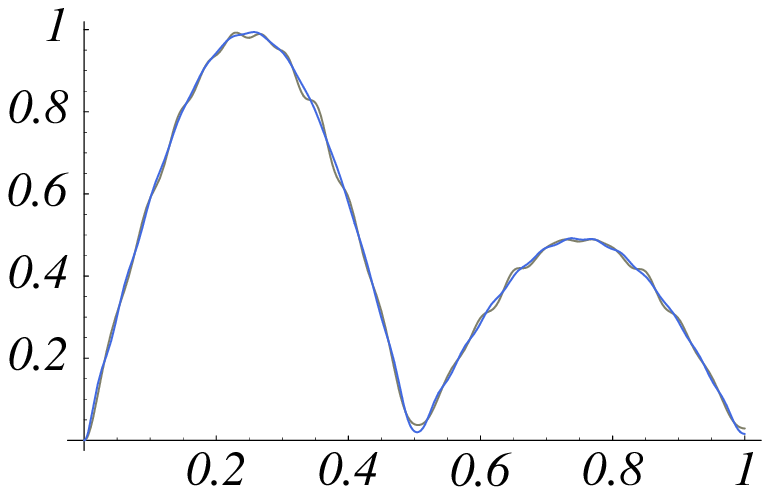}
\includegraphics[scale=.55]{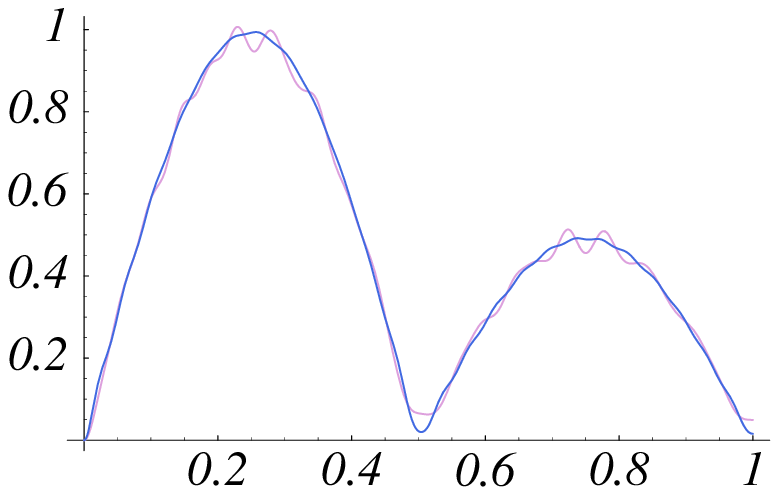}
\end{center}
\caption{\footnotesize {\underline{ $ I(\rho)=
\sin(2\pi \rho)$ for $\rho \in [0,\,0.5)$ and $I(\rho)=\frac{1}{2}
\sin(2\pi \rho)$ for $\rho \in [0.5,\,1]$}}.
The intensity of light is plotted: $\,a)$  shows initial
$I(\rho)$ (an origin and an approximation  given by the
first 50 terms of BF series, plotted together);
$\,b)$ shows the first Talbot revival at $z=z_T$ and the
initial intensity  shown together;
$\,c)$, $\,d)$, $\,e)$ show  2nd, 5th, and 10th Talbot revival,
respectively. }\label{sinn104}
\end{figure}

$\,$\\
To present the example of the revivals of initial intensity,
$I_{org}(0)$, at the Talbot distance and its multiples we have
chosen intensity function of the form:
\be
I_{org}(0)=\left\{\begin{array}{lcl} \sin 2\pi \rho &
\mathrm{for} &\rho\in[0,\,0.5),\;\; \varphi \in [0,2\pi[\\
                            \frac{1}{2}\sin 2\pi \rho &
\mathrm{for} &\rho\in(0.5, 1],\;\;\, \varphi \in [0,2\pi[ .
        \end{array}\right. \n \label{ini}
\ee
Figures
\ref{sinn},
\ref{sinn104} show cross-sections for arbitrary $\varphi$ of this initial
light intensity $I(0)$ and its Talbot revivals obtained for
$ak=10^3$ and $ak=10^4$, respectively.

Analytical function from Eq. (\ref{ini}) and its approximation by
Bessel functions corresponding to $TM_{0i}$ modes are plotted
together in Figures
\ref{sinn}.a,
\ref{sinn104}.a. The numerical procedure allowing to approximate
a given intensity function we have used here is described in the
``Methods" section at the end of the paper. Quality of this
approximation, $f_0^a$, is calculated using measure from Eq.
\ref{measure}. To indicate that the infidelity of approximation is measured,
we shall write
$f_0^a$,
while the infidelities of revivals defined in a similar way are
denoted without this `` $^a$ " superscript.

Figures \ref{sinn}.b,
\ref{sinn104}.b present the first Talbot revival of the initial
fields, Figures \ref{sinn}.c,
\ref{sinn104}.c;
\ref{sinn}.d,
\ref{sinn104}.d; \ref{sinn}.e, \ref{sinn104}.e show 2nd, 5th, and 10th Talbot revivals,
respectively. Above every plot the corresponding infidelities are
depicted. It is very interesting to compare plots and infidelities
obtained for the same field distribution, but for different values
of $ak$, as it is clearly seen that infidelities are much lower
for smaller
$\lambda$ to
$a$ ratio. This effect is connected with the quality of
linear approximation of the square root from Eq.
(\ref{przyblizenie}). The smaller percentage of all modes is used
to construct the initial picture the lower infidelities of higher
revivals are to be expected.

\subsubsection{Some comments on approximations used}
Revivals presented in previous paragraph are quite faithful which
means that approximations used to predict the existence of
revivals were justified. However, please note, that the examples
studied in the previous subsection had the following property:
superposed modes were of the same angular dependence (they have
corresponded to the Bessel functions of fixed
\mbox{number $\nu$)}. Studying more complicated combinations one
finds out that superpositions of modes with different azimuthal
mode numbers do not revive at the Talbot distance \cite{thesis}.

Numerical simulations of field propagation show that the situation
is really
interesting. As we have already mentioned, 
superposition of TM or TE modes having the same angular dependance
revive quite faithfully at Talbot distance and only for the
superpositions of ``mismatched" modes revivals are not obtained.
Explanation of this surprising fact is the following:
the higher terms of the asymptotic formulae for the roots of
Bessel functions, and its derivatives 
\cite{olver}: {\small
\ba
j_{\nu n}=
\underbrace{n\pi+\bigg(\nu-\frac{1}{2}\bigg)\frac{\pi}{2}}_A-
\underbrace{\frac{4\nu^2-1}{8\left(n\pi+\big(\nu-\frac{1}{2}\big)\frac{\pi}{2}\right)}}_B-
\underbrace{\frac{(4\nu^2-1)(28\nu^2-31)}{384\left(n\pi+
\big(\nu-\frac{1}{2}\big)\frac{\pi}{2}\right)^3}}_C-...
\label{jl} \ea }
and {\small
\ba
j'_{\nu n}=
\underbrace{n\pi+\bigg(\nu-\frac{3}{2}\bigg)\frac{\pi}{2}}_{A'}-
\underbrace{
\frac{4\nu^2+3}{8\left(n\pi+\big(\nu-\frac{3}{2}\big)\frac{\pi}{2}\right)}}_{B'}-
\underbrace{
\frac{112\nu^4+328\nu^2-9}{384\left(n\pi+\big(\nu-\frac{3}{2}\big)\frac{\pi}{2}
\right)^3}}_{C'}-...
\label{j'l}
\ea
}are more important then we have assumed so far. In many papers
and textbooks only first terms ($A,\, A'$) of above approximations
are used and we have also limited ourselves to them in preliminary
calculations but to obtain a faithful approximation at least  two
first terms should be taken into account. From formulas
(\ref{jl}), (\ref{j'l}) it is, however, clear that when higher
terms are taken into account, finding  the Talbot distance for
arbitrary $\nu$ and
$n$ fails, because different powers of $\pi$ would be included.

\begin{figure}[h]
\begin{center}
$\!\!$\includegraphics[scale=.7]{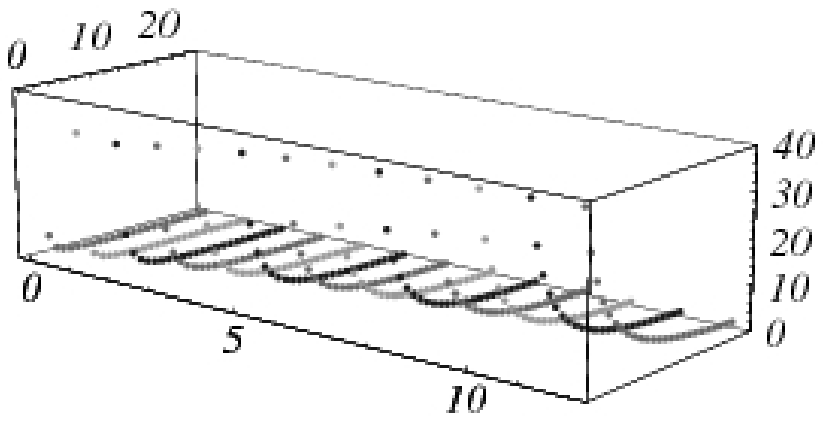}$ \quad$
\includegraphics[scale=.7]{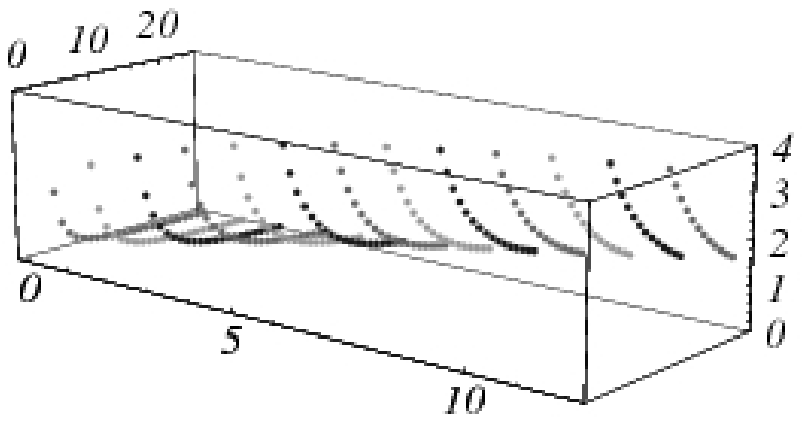}\\
$\!\!$\includegraphics[scale=.7]{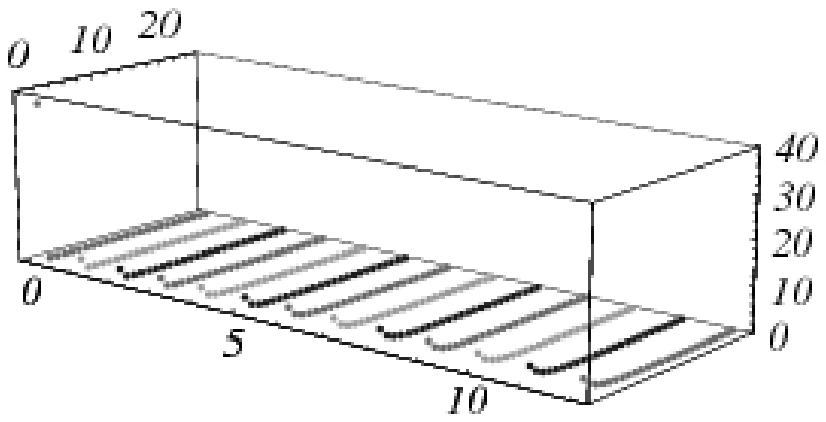}$ \quad$
\includegraphics[scale=.7]{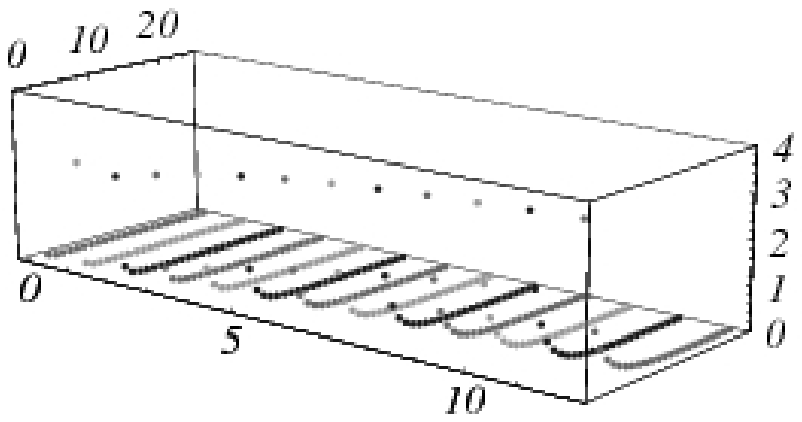}\\
\end{center}
\begin{picture}(0,0)(35,10)
\put(40,118){\makebox(0,0){$a)$}}
\put(2405,118){\makebox(0,0){$b)$}}
\put(40,30){\makebox(0,0){$c)$}}
\put(245,30){\makebox(0,0){$d)$}}
\put(170,110){\makebox(0,0){\footnotesize $\nu$}}
\put(92,195){\makebox(0,0){\footnotesize $n$}}
\put(215,185){\makebox(0,0){\scriptsize $\frac{2B'}{A'}$\%}}
\put(215,98){\makebox(0,0){\scriptsize $\frac{{B'}^2}{{A'}^2}$\%}}
\put(170,24){\makebox(0,0){\footnotesize $\nu$}}
\put(93,108){\makebox(0,0){\footnotesize $n$}}
\put(370,111){\makebox(0,0){\footnotesize $\nu$}}
\put(290,194){\makebox(0,0){\footnotesize $n$}}
\put(408,185){\makebox(0,0){\scriptsize   $\frac{2B'}{A'}$\%}}
\put(412,95){\makebox(0,0){\scriptsize  $\frac{{B'}^2}{{A'}^2}$\%}}
\put(368,23){\makebox(0,0){\footnotesize $\nu$}}
\put(288,108){\makebox(0,0){\footnotesize $n$}}
\end{picture}
\caption{\footnotesize {\sl\underline{Errors of ``asymptotic"
approximation of $(j'_{\nu n})^2$}}: Notation is taken from from
Eqs. \ref{j'l},
\ref{eq12}: Plots a) and b) show in two different scales percentage ratio
of $2B'/A'$ for $\nu\in\{0,...,12\}$, $n\in\{1,...,25\} $.
Unicolor dotted lines correspond to one value of azimuthal mode
number $\nu$. Plots c) and d) present percentage ratio of
${B'}^2/{A'}^2$. It is seen that for $n>\nu$ this term can be
neglected. }
\label{blad_a}
\end{figure}

So how we can observe any revivals at all? Let us take a closer
look at approximation (\ref{j'l}), i.e. one considering TE modes
-- keeping in mind that similar analysis can be made for TM modes
as well. Taking two first terms of approximation (\ref{j'l}) the
following expression for
${j'}_{\nu n}^2$ is obtained:
\ba
{j'}_{\nu
n}^2=\underbrace{\left(n\pi+\bigg(\nu-\frac{3}{2}\bigg)
\frac{\pi}{2}\right)^2}_{{A'}^2}-
\underbrace{
\frac{4\nu^2+3}{4}}_{2 A'B'}+
\underbrace{
\frac{(4\nu^2+3)^2}{64\left(n\pi+
\big(\nu-\frac{3}{2}\big)\frac{\pi}{2}\right)^2}}_{{B'}^2}
\label{eq12}
\ea
Ratios ${B'}^2/{A'}^2$ and $2 A' B'/{A'}^2$ for a wide range of
parameters $\nu$ and $n$ are shown in Figure
\ref{blad_a}. It is seen that although the ratio  ${B'}^2/{A'}^2$ for
$n>\nu$  is close to zero and can be neglected, the percentage
value of $2 B'/A'$ can be quite significant.

Obviously, this is the effect we were looking for as term $2
A'B'=\nu^2+\frac{3}{4}$ depends only on the azimuthal mode number
$\nu$ and not on the radial one $n$. That is the reason why modes having
the same angular dependence revive at the Talbot distance
$z_T=\frac{16}{\lambda}$, while for a superposition of modes with
different angular dependence we do not obtain such revivals. In
the first case factor $
\exp\left(-i(4\nu^2+3)z_T/4\right)$ is merely a phase common for
all the terms of the superposition. When the initial $H_z$ is the
superposition of a form $H_z(0)=J_\nu(j'_{\nu
n}\rho)e^{i\nu\varphi}+ J_\mu(j'_{\mu m}\rho)e^{i\mu\varphi}   \;
$ then at $z=z_T$ phase factors corresponding to $J_\nu$ and
$J_\mu$ differ for $\nu\neq\mu$ and, consequently, they do not
cancel. This relative phase destroys the Talbot revivals promised
by less accurate approximation.

Similar analysis shows that superpositions of $TM$ and $TE$ modes
will not revive at the same distance because of the relative
phase. Only for $TM_{1i}$ and $TE_{0j}$ modes this relative phase
disappears and Talbot effect can be observed. However, due to
possibility of using polarizers, restrictions imposed by necessity
of choosing polarization of modes used is experimentally less
demanding then that concerning
$\varphi$-dependance.


\section{Summary}
We have discussed in details approximations that appear in the
study of 
the Talbot effect in the cylindrical mirror waveguides as well as
different
 approximations used
in the case of dielectric waveguides. 
We have shown that in many cases almost perfect revivals can be
obtained and that even dephased propagation can be used in
practice. 
We have stressed that {\mbox{a phase}} space description
(sometimes regarded as an unnecessary complicated representation)
extracts an essence of the interference phenomena and that the
conditions needed to be fulfilled to obtain Talbot revivals in a
very natural way  follow from a phase space description of
interference.
{\footnotesize
\section{Methods}
The following numerical procedure was used for decomposing given
initial light intensity $I_{org}$ into $TM_{0i}$ modes: For an
arbitrary spherically symmetric $TM$ field with an initial $E_z$
of a form
\ba
E_z(0)=\sum_{i=1}^{50} x_i J_0(j_{0i}\,\rho)
\ea
we have evaluated the light intensity $I(0)$ and this $I(0)$ was
expanded again in a Bessel-Fourier series corresponding to
$J_0$. Numerical evaluation of some integrals was required at this
point. In such a way  general ``basis" $(e_1,e_2,...,e_{50})$ was
obtained, (every
$e_i$ being a sum of all possible pairs $x_i
\cdot  x_j$ with numerically calculated coefficients). Then, an
arbitrary $\varphi$-independent intensity $I_{org}$ that we would
like to propagate through the waveguide was decomposed in {\sl BF}
series
\ba
I_{org}(0) \approx \sum_{i=1}^{50} c_i J_0(j_{0i}\rho),
\ea
and a set of quadratic equations $c_1=e_1$, $c_2=e_2,\,\cdots\,,$
$c_{50}=e_{50}$ was solved numerically to find $(x_i)_{i=1}^{50}$.
Figures \ref{sinn}.a, \ref{sinn104}.a  can be treated as a test of
faithfulness of the solutions founded in the procedure described
above -- the infidelities of approximation $f_0^a$ are of the
order of $ 10^{ - 2}$. }

\section{Acknowledgements}
We would like to acknowledge useful discussions with Professor W.
P. Schleich. This research was partially supported by Polish MEN
Grant No. 1 P03B 137 30 and European Union's Transfer of Knowledge
project CAMEL (Grant No. MTKD-CT-2004-014427).

\end{document}